\pdfoutput=1
\RequirePackage{ifpdf}
\ifpdf 
\documentclass[pdftex]{sigma}
\else
\documentclass{sigma}
\fi

\usepackage{cite}

\numberwithin{equation}{section}
\numberwithin{theorem}{section}
\numberwithin{proposition}{section}
\numberwithin{lemma}{section}
\numberwithin{corollary}{section}
\numberwithin{definition}{section}
\numberwithin{example}{section}
\numberwithin{remark}{section}

\newcommand{\cA}{\mathcal{A}}
\newcommand{\cB}{\mathcal{B}}

\newcommand{\cP}{\mathcal{P}}
\newcommand{\cR}{\mathcal{R}}
\newcommand{\cU}{\mathcal{U}}
\newcommand{\cV}{\mathcal{V}}

\newcommand{\bA}{\boldsymbol{A}}

\newcommand{\bGamma}{\boldsymbol{\Gamma}}
\newcommand{\bI}{\boldsymbol{I}}

\newcommand{\bN}{\boldsymbol{N}}
\newcommand{\bP}{\boldsymbol{P}}

\newcommand{\bQ}{\boldsymbol{Q}}

\newcommand{\bU}{\boldsymbol{U}}
\newcommand{\bV}{\boldsymbol{V}}
\newcommand{\bW}{\boldsymbol{W}}
\newcommand{\bX}{\boldsymbol{X}}
\newcommand{\bY}{\boldsymbol{Y}}

\newcommand{\bbA}{\mathbb{A}}
\newcommand{\bbC}{\mathbb{C}}

\newcommand{\bbR}{\mathbb{R}}

\newcommand{\tphi}{\tilde{\phi}}

\newcommand{\tbP}{\tilde{\boldsymbol{P}}}
\newcommand{\tbQ}{\tilde{\boldsymbol{Q}}}
\newcommand{\tbX}{\tilde{\boldsymbol{X}}}

\renewcommand{\d}{\mathrm{d}}
\newcommand{\bd}{\bar{\mathrm{d}}}

\newcommand{\imag}{\mathrm{i}}

\newcommand{\zz}{z}

\begin{document}

\allowdisplaybreaks

\renewcommand{\PaperNumber}{009}

\FirstPageHeading

\ShortArticleName{Binary Darboux Transformations and Vacuum Einstein Equations}

\ArticleName{Binary Darboux Transformations in Bidif\/ferential\\ Calculus
 and Integrable Reductions\\ of Vacuum Einstein Equations}

\Author{Aristophanes DIMAKIS~$^\dag$ and Folkert M\"ULLER-HOISSEN~$^\ddag$}

\AuthorNameForHeading{A.~Dimakis and F.~M\"uller-Hoissen}

\Address{$^\dag$~Department of Financial and Management Engineering,
 University of the Aegean,\\
 \hphantom{$^\dag$}~82100 Chios, Greece}
\EmailD{\href{mailto:dimakis@aegean.gr}{dimakis@aegean.gr}}

\Address{$^\ddag$~Max-Planck-Institute for Dynamics and Self-Organization,
         37077 G\"ottingen, Germany}
\EmailD{\href{mailto:folkert.mueller-hoissen@ds.mpg.de}{folkert.mueller-hoissen@ds.mpg.de}}

\ArticleDates{Received November 12, 2012, in f\/inal form January 29, 2013; Published online February 02, 2013}

\Abstract{We present a general solution-generating result within the bidif\/ferential
calculus approach to integrable partial dif\/ferential and dif\/ference equations, based on a
binary Darboux-type transformation. This is then applied to the non-autonomous chiral model,
a certain reduction of which is known to appear in the case of the $D$-dimensional vacuum
Einstein equations with $D-2$ commuting Killing vector f\/ields.
A large class of exact solutions is obtained, and the aforementioned reduction is
implemented. This results in an alternative to the well-known Belinski--Zakharov formalism.
We recover relevant examples of space-times in dimensions four (Kerr-NUT,
Tomimatsu--Sato) and f\/ive (single and double Myers--Perry black holes, black saturn,
bicycling black rings).}


\Classification{37K10; 16E45}

\section{Introduction}
The bidif\/ferential calculus formalism aims to understand integrability features
and solution-generating methods for (at least a large class of) \emph{integrable}
partial dif\/ferential or dif\/ference equations (PDDEs) resolved from the particularities
of examples,
i.e., on an as far as possible \emph{universal} level \cite{DMH00a,DMH00e,DMH08bidiff}.
The most basic ingredient is a graded associative algebra, supplied with two
anti-commuting graded derivations of degree one.
It can and should be regarded as a~generalization (in the spirit of noncommutative geometry)
of the algebra of dif\/ferential forms on a~manifold, but supplied with \emph{two}
analogs of the exterior derivative. Once a PDDE is translated to this framework,
it is simple to elaborate its integrability conditions. In fact, it could not
be simpler.

In this framework, Darboux transformations
(see~\cite{Matv+Sall91,Roge+Schi02,Gu+Zhou05} and the references therein) have f\/irst been addressed
in~\cite{DMH08bidiff}. In the latter work, we had also obtained a very simple solution
gene\-ra\-ting result that evolved into improved versions in recent applications
\cite{DMH10NLS,DMH10AKNS,DKMH11sigma}. Its relation with Darboux transformations has been
further clarif\/ied in~\cite{DKMH11sigma} (see Appendix~A therein), an essential step toward
the much more general result that we present in this work. The resulting class of solutions
is expressed in a \emph{universal} way, in the sense that the corresponding formula holds
simultaneously for all integrable PDDEs possessing a bidif\/ferential calculus formulation.
\mbox{Choosing} a bidif\/ferential calculus associated with a specif\/ic PDDE, we can generate
inf\/inite families of (soliton-like) solutions.

In Section~\ref{sec:Darboux} we recall some basics of bidif\/ferential calculus and set
up the stage in a concise way to formulate our general result about Darboux transformations.
This hardly requires previous knowledge and can be taken as an independent and self-contained
step into the world of integrable PDDEs. That this is indeed a very powerful tool, is
demonstrated in the rest of this work, where we concentrate on one of the more tricky
examples of integrable PDEs.

In Section~\ref{sec:naCM} we elaborate in detail the example of the non-autonomous
chiral model equation
\begin{gather}
   \big(\rho   g_\zz   g^{-1}\big)_\zz + \epsilon   \big(\rho   g_\rho   g^{-1}\big)_\rho = 0  \label{naCM}
\end{gather}
for an $m \times m$ matrix $g$, where $\rho>0$ and $\zz$ are independent real variables and
$\epsilon = \pm 1$.\footnote{A subscript indicates a partial derivative with respect to the
corresponding variable. Formally, $\rho \mapsto \imag \rho$ relates the two values of
$\epsilon$ in~(\ref{naCM}).}
With $\epsilon=1$, this governs
the case of the stationary, axially symmetric vacuum Einstein  ($m=2$) and Einstein--Maxwell ($m=3$)
equations in four dimensions, where we have two commuting Killing vector f\/ields,
one spacelike and the other
one asymptotically timelike (see, e.g., \cite{Beli+Sakh79,Beli+Verd01,SKMHH03,Klein+Richter05}).
In an analogous way, (\ref{naCM}) with $m>3$ appears in the dimensional reduction of
(the bosonic part of) higher-dimensional supergravity theories to two dimensions
(see, e.g., \cite{BMG88,Harm04,Empa+Real08,FJRV10,Tomi+Ishi11,IIM11}).
With $\epsilon=-1$, the above equation appears in the case of vacuum solutions of
Einstein's equations in four dimensions with two commuting spacelike Killing vector f\/ields
\cite{Beli+Zakh78,Griff91,Verd93,Beli+Verd01}, describing in particular cylindrical gravitational waves.

In Section~\ref{sec:naCM}, we also present a reduction condition that, imposed on the
obtained family of solutions, achieves that $g$ is \emph{symmetric}.
Any such real and symmetric~$g$ determines a solution of the vacuum Einstein equations in $m+2$
dimensions, as recalled in Section~\ref{sec:Einstein}. The resulting recipe to construct
solutions of the vacuum Einstein equations is close to the familiar Belinski--Zakharov
method \cite{Beli+Zakh78,Beli+Sakh79,Beli+Verd01}, which has been applied in numerous
publications. In a sense, what we obtained is a kind of matrix version of the latter.

A well-known reformulation of the equations obtained from the integrable reduction of the four-dimensional
vacuum Einstein (and also the Einstein--Maxwell) equations by introduction of the so-called
twist potential (a crucial step toward the Ernst equation), connects space-time metrics in a dif\/ferent way
with the non-autonomous chiral model (see, e.g., \cite{Naka83}). Solutions of the latter are then required to have
a \emph{constant} determinant. This allows a \emph{constant} seed and thus an application
of the restricted solution-generating result in \cite{DKMH11sigma} (also see Remark~\ref{rem:DKM11} below).
It does not work in the more direct approach we take in the present work, but here we resolve the restriction
in~\cite{DKMH11sigma}.

In Section~\ref{sec:Einstein} we elaborate several examples in
four and f\/ive space-time dimensions, and show that the resulting metrics include
important solutions of Einstein's equations, in particular some of the more recently found
black objects in f\/ive dimensions, which have no counterpart in four dimensions
(see, e.g., \cite{Empa+Real08,Tomi+Ishi11,IIM11,Horowitz12}).
Here we used {\sc Mathematica}\footnote{Mathematica Edition: Version 8, Wolfram Research, Inc., Champaign, Illinois,
  2010.} 
  in a substantial way.

Section~\ref{sec:conclusions} contains some concluding remarks.

\section{Binary Darboux transformations in bidif\/ferential calculus}
\label{sec:Darboux}
A \emph{graded associative algebra} is an associative algebra $\Omega$ over $\bbC$ with a direct
sum decomposition $\Omega = \bigoplus_{r \geq 0} \Omega^r$
into a subalgebra $\cA := \Omega^0$ and $\cA$-bimodules $\Omega^r$, such that
$\Omega^r   \Omega^s \subseteq \Omega^{r+s}$.
A~\emph{bidifferential calculus} (or \emph{bidifferential graded algebra}) is a unital
graded associative algebra~$\Omega$, equipped with two ($\bbC$-linear) graded
derivations\footnote{Hence $\d$ and $\bd$ both satisfy the graded Leibniz rule
$
    \d(\chi   \chi') = (\d \chi)   \chi' + (-1)^r   \chi   \d \chi'    ,
$
for all $\chi \in \Omega^r$ and $\chi' \in \Omega$.}
$\d, \bar{\d} : \Omega \rightarrow \Omega$ of degree one (hence $\d \Omega^r \subseteq \Omega^{r+1}$,
$\bar{\d} \Omega^r \subseteq \Omega^{r+1}$), with the properties
\begin{gather}
    \d^2 = \bd^2 = \d \bd + \bd \d = 0    .   \label{bidiff_conds}
\end{gather}

In the bidif\/ferential calculus approach to integrable PDDEs we are looking for a choice
$(\Omega, \d, \bd)$ and some $\phi \in \cA$, or some invertible $g \in \cA$,
such that either
\begin{gather}
     \bd   \d   \phi = \d\phi   \d\phi     \label{phi_eq}
\end{gather}
or
\begin{gather}
        \d \big[ (\bd g)   g^{-1} \big] = 0     \label{g_eq}
\end{gather}
is equivalent to a certain PDDE. The two equations are related by
the \emph{Miura equation}
\begin{gather}
     (\bd g)   g^{-1} = \d \phi   ,  \label{Miura}
\end{gather}
which establishes a kind of \emph{Miura transformation}\footnote{The original \emph{Miura transformation}
maps solutions of the modif\/ied Korteweg--de~Vries equation to solutions of the Korteweg--de~Vries equation.}
between the two equations~(\ref{phi_eq}) and~(\ref{g_eq}).
This equation has both, (\ref{phi_eq}) and~(\ref{g_eq}), as integrability conditions.
As a consequence, if we f\/ind a solution pair $(\phi,g)$
of~(\ref{Miura}), then $\phi$ solves~(\ref{phi_eq}) and~$g$ solves~(\ref{g_eq}).

In the following theorem, we formulate a solution-generating result which amounts
to a~transformation that takes a given solution $(\phi_0,g_0)$ of the Miura equation
(\ref{Miura}) to a new solution of~(\ref{Miura}),
\[
      (\phi_0, g_0) \quad \mapsto \quad  (\phi, g)    .
\]
This induces corresponding transformations of solutions of (\ref{phi_eq}), respectively (\ref{g_eq}).

Let now $\cA$ be the algebra of all f\/inite-dimensional\footnote{An extension to a suitable
class of $\infty$-dimensional matrices, respectively operators, is certainly possible.}
matrices, with entries in a unital algebra~$\cB$. The product of two matrices is
def\/ined to be zero if the sizes of the two matrices do not match. In the following
we assume that there is a graded algebra $\Omega$ with $\Omega^0 = \cA$, and a bidif\/ferential
calculus $(\Omega,\d, \bd)$, and such that $\d$ and $\bd$ preserve the size of matrices.
$I=I_m$ and $\bI=\bI_n$ denote the $m \times m$, respectively
$n \times n$, identity matrix, and we assume that they are annihilated by~$\d$ and~$\bd$.
Furthermore, $\mathrm{Mat}(m,n,\cB)$ denotes the set of $m \times n$ matrices
over~$\cB$.

\begin{theorem}\sloppy
\label{thm:main}
Let $\phi_0, g_0 \in \mathrm{Mat}(m,m,\cB)$ solve the Miura equation \eqref{Miura}.
Let $\bP,\bQ \in \mathrm{Mat}(n,n,\cB)$ be invertible solutions of
\begin{gather}
   \bar{\d} \bP = (\d \bP)   \bP   , \qquad
   \bd \bQ = \bQ   \d \bQ   ,   \label{P,Q_eqs}
\end{gather}
which are \emph{independent} in the sense that $\bQ   \bY = \bY   \bP$ implies $\bY=0$.
Let $\bU \in \mathrm{Mat}(m,n,\cB)$ and $\bV \in \mathrm{Mat}(n,m,\cB)$ be solutions
of the linear equations\footnote{If we are primarily interested in solving~(\ref{g_eq}),
it is more convenient to replace $\d \phi_0$ by $(\bd g_0)  g_0^{-1}$ in these equations,
by use of the Miura equation for $(\phi_0,g_0)$. }
\begin{gather}
    \bd \bU = (\d \bU)   \bP + (\d \phi_0)   \bU   , \qquad
    \bd \bV = \bQ   \d \bV - \bV   \d \phi_0   .    \label{U,V_eqs}
\end{gather}
Furthermore, let $\bX \in \mathrm{Mat}(n,n,\cB)$ be an invertible solution of the Sylvester-type
equation
\begin{gather}
   \bX   \bP - \bQ   \bX = \bV   \bU   .  \label{preSylvester}
\end{gather}
Then
\begin{gather}
    \phi = \phi_0 + \bU \bX^{-1} \bV   , \qquad
       g = \big(I + \bU (\bQ   \bX)^{-1} \bV\big)   g_0    \label{thm_phi,g}
\end{gather}
solve the Miura equation~\eqref{Miura}, and thus also \eqref{phi_eq},
respectively~\eqref{g_eq}.
\end{theorem}

\begin{proof}
Acting on (\ref{preSylvester}) with $\bd$, using the rules of
bidif\/ferential calculus, (\ref{P,Q_eqs}), (\ref{U,V_eqs}), and (\ref{preSylvester}) again,
we f\/ind
\[
  \bQ   [ \bd \bX - (\d \bX)   \bP + (\d \bQ)   \bX + (\d \bV)   \bU ] =
     [ \bd \bX - (\d \bX)   \bP + (\d \bQ)   \bX + (\d \bV)   \bU ]   \bP  .
\]
Since $\bP$ and $\bQ$ are assumed to be independent, this implies\footnote{It is a key feature
of Theorem~\ref{thm:main} that this equation is automatically solved if
$\bP$ and $\bQ$ are independent. Relaxing the latter assumption, the theorem remains valid
if this equation is added to the assumptions.}
\begin{gather}
    \bd \bX - (\d \bX)   \bP + (\d \bQ)   \bX + (\d \bV)   \bU  = 0   .
   \label{inhomog_ext_lin_sys}
\end{gather}
Using also (\ref{P,Q_eqs}) and (\ref{preSylvester}), we easily deduce that
\begin{gather*}
      \bd (\bQ \bX)^{-1}
   =  - \bX^{-1} \big[ (\d \bX)   \bX^{-1} (\bX \bP)  - (\d \bV) \bU \big]   (\bQ \bX)^{-1} \\
\hphantom{\bd (\bQ \bX)^{-1}}{}
   =  (\d \bX^{-1})   \big[ \bI + \bV   \bU   (\bQ \bX)^{-1}\big]
      + \bX^{-1} (\d \bV)   \bU (\bQ \bX)^{-1}   .
\end{gather*}
This is then used in the elaboration of
\[
    \bd g = \bd \big[ (I + \bU (\bQ \bX)^{-1} \bV)   g_0 \big]
\]
via the graded derivation (Leibniz) rule for $\bd$. We eliminate all further terms involving a~$\bd$
with the help of~(\ref{U,V_eqs}), and apply~(\ref{preSylvester}) to eliminate a $\bP$ in favor of a~$\bQ$.
Finally we obtain $\bd g = (\d \phi)   g$ with $\phi$ given by~(\ref{thm_phi,g}).
Our assumptions ensure that the inverse of $g$ exists. It is given by
\[
     g^{-1} = g_0^{-1} \big(I - \bU (\bX \bP)^{-1} \bV\big)   .
\]
The integrability conditions of (\ref{U,V_eqs}) and (\ref{inhomog_ext_lin_sys}) are satisf\/ied
as a consequence of~(\ref{P,Q_eqs}) and $\bd \d \phi_0 = \d \phi_0   \d \phi_0$ (which
follows from $(\bd g_0)   g_0^{-1} = \d \phi_0$).
\end{proof}

For a reader acquainted with corresponding results about binary Darboux transformations in
the literature on integrable systems, the similarity will be evident (see, e.g.,
\cite{Matv+Sall91,Roge+Schi02,Gu+Zhou05,NGO00,Cies09}).
It is close to results in \cite{Sakh94,Sakh01,Sakh10}\footnote{The precise relation has still
to be clarif\/ied.}.
A new feature, however, is the dependence of the linear equations~(\ref{U,V_eqs}) on
solutions~$\bP$ and $\bQ$ of the \emph{nonlinear} equations~(\ref{P,Q_eqs}). This
generalization is crucial for our application to the Einstein equations.
As already mentioned in the introduction, Theorem~\ref{thm:main} considerably generalizes previous,
more restricted results, see in particular \cite{DKMH11sigma}.

\begin{remark}
As a consequence of~(\ref{P,Q_eqs}), $\bP$ and $-\bQ$ solve the $n \times n$ version of (\ref{phi_eq}).
Since we assume that they are invertible, $\bP$ and $\bQ^{-1}$ solve the
$n \times n$ version of~(\ref{g_eq}).
If $\bU$ is a solution of the f\/irst of~(\ref{U,V_eqs}), then also $\bU \bP$, by use of the
f\/irst of (\ref{P,Q_eqs}). If $\bV$ is a solution of the second of~(\ref{U,V_eqs}), then
also $\bQ \bV$, by use of the second of~(\ref{P,Q_eqs}).
\end{remark}

\begin{remark}
\label{rem:scaling}
By direct computations one verif\/ies the following ``scaling property''.
Let~$\bU$ and~$\bV$ solve (\ref{U,V_eqs}) and let
\[
    \hat{g}_0 = w   g_0   , \qquad
    \hat{\bU} = \bU   \bW_1   , \qquad
    \hat{\bV} = \bW_2   \bV   ,
\]
with a solution $(\varphi,w)$ of the $m=1$ version of (\ref{Miura}),
and invertible $\bW_i \in \mathrm{Mat}(n,n,\cB)$, subject to
\[
    [\bW_1, \bP] = 0   , \qquad [\bW_2, \bQ] = 0   .
\]
We further assume that $\varphi$ and $w$ are in the center of $\Omega$.\footnote{This
holds, e.g., in the case of the non-autonomous chiral model treated in Section~\ref{sec:naCM}.
But this assumption will be a (perhaps too) severe restriction if $\cB$ involves dif\/ferential
or dif\/ference operators.}
If $\bW_1$ and $\bW_2$ satisfy
\[
    \bd \bW_1 = (\d \bW_1)   \bP + (\bd w)   w^{-1}   \bW_1   , \qquad
    \bd \bW_2 = \bQ  \d \bW_2 - \bW_2   (\bd w)   w^{-1}    ,
\]
then $\hat{\bU}$ and $\hat{\bV}$ solve (\ref{U,V_eqs}), with $\phi_0$
replaced by $\hat{\phi}_0 = \phi_0 + \varphi   I$.
Furthermore, $\hat{\bX} = \bW_2 \bX \bW_1$ solves (\ref{preSylvester})
with $\bU$, $\bV$ replaced by $\hat{\bU}$, $\hat{\bV}$. For the new solutions of~(\ref{Miura}) (and thus~(\ref{phi_eq}), respectively~(\ref{g_eq})), determined by
the theorem, we f\/ind
\[
    \hat{\phi} = \phi + \varphi   I   , \qquad
       \hat{g} = w   g   ,
\]
where $\phi$ and $g$ are given by~(\ref{thm_phi,g}).
Hence, under the stated conditions, multiplication of $g_0$ by a ``scalar'' solution~$w$
of~(\ref{g_eq}) simplify amounts to multiplication of~$g$ by~$w$.
In the special case where $w$ is the identity element and $g$ thus remains unchanged,
the above equations for~$\bW_1$ and~$\bW_2$ still allow non-trivial transformations of~$\bU$ and~$\bV$.
These results will be used in Section~\ref{sec:Einstein}.
\end{remark}

\begin{remark}
\label{rem:superpos}
Let $(\phi_i, g_i)$, $i=1,2$, be two solutions obtained via Theorem~\ref{thm:main} from the \emph{same}
seed solution $(\phi_0, g_0)$. Let $(\bP_i, \bQ_i, \bU_i, \bV_i)$ be corresponding solutions
of (\ref{P,Q_eqs}) and (\ref{U,V_eqs}). Then
\[
    \bP = \left( \begin{matrix} \bP_1 & 0 \\ 0 & \bP_2
                  \end{matrix} \right)   , \qquad
    \bQ = \left( \begin{matrix} \bQ_1 & 0 \\ 0 & \bQ_2
                  \end{matrix} \right)   , \qquad
    \bU = (\bU_1, \bU_2)   , \qquad
    \bV = \left( \begin{matrix} \bV_1 \\ \bV_2 \end{matrix} \right)
\]
also solve~(\ref{P,Q_eqs}) and~(\ref{U,V_eqs}). If $\bP$ and $\bQ$ are independent and if~(\ref{preSylvester}) has an invertible solution $\bX$, then (\ref{thm_phi,g}) determines a~new solution of~(\ref{Miura}), and thus new solutions of~(\ref{phi_eq}) and~(\ref{g_eq}).
This expresses a \emph{nonlinear superposition principle}. By iteration, one can build superpositions
of an arbitrary number of ``elementary'' solutions, hence (analogs of) ``multi-solitons''.
\end{remark}

\begin{remark}
We note that the f\/irst of equations (\ref{U,V_eqs}) is a special case (and $n \times n$ matrix version)
of the general \emph{linear equation}
\begin{gather}
    \bd \psi - \bbA   \psi = \d \psi   P,  \qquad \mbox{where} \quad \bbA = \d \phi   .
            \label{bidiff_lin_eq}
\end{gather}
Here $P$ is a solution of the nonlinear equation
\[
     \bd P = (\d P)   P    .
\]
An $n \times n$ matrix version of it appears in~(\ref{P,Q_eqs}).
The integrability condition of (\ref{bidiff_lin_eq}) is~(\ref{phi_eq}).
If we set $\bbA = (\bd g)   g^{-1}$, the integrability condition is~(\ref{g_eq}) instead.
A similar statement holds for the second of equations (\ref{U,V_eqs}), the ``adjoint linear system''.
The Darboux transformation (\ref{thm_phi,g}) is ``binary'' since it involves solutions
of the linear system as well as of the adjoint.
\end{remark}

\begin{remark}
In the special case where $\Omega$ is the algebra of matrix-valued dif\/ferential forms on a manifold,
and $\d$ the exterior derivative, any tensor f\/ield $\mathcal{N}$ of type $(1,1)$ with
vanishing Nijenhuis torsion determines a map $\bd_{\mathcal{N}}$ satisfying the above
conditions \cite{Froe+Nije56,Chav05}. Moreover, according to Fr\"olicher--Nijenhuis theory
\cite{Froe+Nije56}, any $\bd$ such that (\ref{bidiff_conds}) holds has to be of this form.
Finite-dimensional integrable systems have been
considered in this framework in~\cite{CST00b}. The ge\-neralization to Lie algebroid
structures is nicely described in~\cite{Cama+Cari03}.
See also~\cite{Lore+Magri05,Lorenzoni06,Arsi+Lore11} for related aspects.
We should stress, however, that in our central examples we depart from
dif\/ferential geometry and consider a dif\/ferential calculus in a weaker sense (e.g., of
noncommutative geo\-metry).
\end{remark}

The setting of the above theorem allows in principle structures far away from
classical calculus and in particular dif\/ferential geometry, then dealing with equations beyond
dif\/ferential and dif\/ference equations.
A somewhat more restricted framework is given by setting
\[
   \Omega = \cA \otimes \bigwedge\big(\bbC^N\big)   ,
\]
where $\bigwedge(\bbC^N)$ denotes the exterior
(Grassmann) algebra of the vector space $\bbC^N$. In this case it is suf\/f\/icient to
def\/ine suitable operators $\d$ and $\bd$ on $\cA$, since they extend to $\Omega$
in an evident way. Many integrable PDDEs have been treated in this framework
and in the next section we turn to an important example.

\section{Solutions of the non-autonomous chiral model}
\label{sec:naCM}

The non-autonomous chiral model is well-known to be a reduction of the (anti-) self-dual
Yang--Mills equations, for which a very simple bidif\/ferential calculus exists~\cite{DMH08bidiff}
(that may actually be considered as a prototype). From the latter
one can then derive a bidif\/ferential calculus for~(\ref{naCM}). It is determined by
\begin{gather*}
    \d f = - f_\zz   \xi_1 + e^\theta   \big(f_\rho - \rho^{-1} f_\theta\big)   \xi_2
             , \qquad
    \bar{\d} f = e^{-\theta}   \big(f_\rho + \rho^{-1} f_\theta\big)   \xi_1 + \epsilon   f_\zz   \xi_2
\end{gather*}
for $f \in C^\infty(\bbR^3)$ (cf.~\cite{DKMH11sigma}).
$\xi_1$, $\xi_2$ is a basis of $\bigwedge^1(\bbC^2)$.
Choosing $\cB = C^\infty(\bbR^3)$,
$\d$ and $\bd$ extend to $\cA \otimes \bigwedge(\bbC^2)$ via $\d (f_1   \xi_1  + f_2   \xi_2)
= (\d f_1) \wedge \xi_1  + (\d f_2) \wedge \xi_2$ and $\d (f   \xi_1 \wedge \xi_2)
  = (\d f) \wedge \xi_1 \wedge \xi_2 =0$, and correspondingly for~$\bd$.
For an $m \times m$ matrix-valued function~$g$, (\ref{g_eq}) now takes the form
\[
    \big(\rho   g_\zz   g^{-1}\big)_\zz + \epsilon   \big(\rho   g_\rho   g^{-1}\big)_\rho
    - \big[\big(g_\rho + \rho^{-1} g_\theta \big)   g^{-1}\big]_\theta + \big(g_\theta   g^{-1}\big)_\rho
     = 0   ,
\]
which reduces to (\ref{naCM}) if $g$ does not depend on $\theta$. The coordinate $\theta$ is
needed to have the properties of a bidif\/ferential calculus, but addressing~(\ref{naCM}) we
are primarily interested in equations for objects that do not depend on it.

Using this bidif\/ferential calculus, the Miura equation (\ref{Miura}) decomposes into
\[
     \big(g_\rho + \rho^{-1} g_\theta\big)   g^{-1} = - e^\theta   \phi_z   , \qquad
     g_z   g^{-1} = \epsilon   e^\theta   \big(\phi_\rho - \rho^{-1} \phi_\theta\big)    .
\]
If $g$ is $\theta$-independent, this requires
\begin{gather}
      \phi = e^{-\theta}   \tphi    ,  \label{naCM_phi->tphi}
\end{gather}
with $\theta$-independent $\tphi$, and then reduces to
\begin{gather}
     g_\rho   g^{-1} = - \tphi_z   , \qquad
     \epsilon   g_z   g^{-1} = \tphi_\rho + \rho^{-1} \tphi   .
          \label{naCM_Miura}
\end{gather}

In the following, we elaborate Theorem~\ref{thm:main} using the above bidif\/ferential
calculus. Section~\ref{subsec:P&Q} provides the complete solution of the two nonlinear
equations~(\ref{P,Q_eqs}) under the condition that $\bP$ (respectively $\bQ$) has
\emph{geometrically simple spectrum} (i.e., for each eigenvalue there is a unique corresponding
Jordan block in the Jordan normal form), also see \cite{DKMH11sigma}.
In this case the two equations actually coincide.
Then it only remains to solve \emph{linear} equations, see Section~\ref{subsec:U&V}
and Section~\ref{subsec:Sylvester} below.
In Section~\ref{subsec:sym_reduction} we present a condition to be imposed on the
data in order to achieve that the solution $g$ is symmetric (or Hermitian).
This reduction is crucial in the context of Einstein's equations, see Section~\ref{sec:Einstein}.
Most of the following is, however, independent of additional assumptions and provides a
general procedure to construct solutions of the non-autonomous chiral model.
The corresponding equations cannot be solved \emph{explicitly} without
some restrictions, in particular on the form of the seed solution.
We content ourselves with providing illustrative and important examples relevant in the
context of gravity in Section~\ref{sec:Einstein}.

\subsection[The equations for $P$ and $Q$]{The equations for $\bP$ and $\bQ$}
\label{subsec:P&Q}
In terms of
\[
    \tbP := e^\theta   \bP   ,
\]
the f\/irst of the two equations (\ref{P,Q_eqs}) decomposes into the following pair of
equations,
\begin{gather}
     \tbP_\rho + \rho^{-1} \big(\tbP_\theta - \tbP\big) = - \tbP_\zz   \tbP
               , \qquad
     \epsilon   \tbP_\zz = \big[\tbP_\rho - \rho^{-1} \big(\tbP_\theta - \tbP\big) \big]   \tbP   ,
        \label{tbP_eqs}
\end{gather}
which are autonomous in the variable $\theta$.
Assuming that $\tbP$ and $\bI + \epsilon   \tbP^2$ are invertible,
and that $\tbP$ is $\theta$-independent, (\ref{tbP_eqs}) implies
\begin{gather}
    \tbP^2 - 2 \rho^{-1}   (\zz   \bI - \bA)   \tbP - \epsilon   \bI = 0   ,
    \label{tbP_id}
\end{gather}
with an arbitrary constant $n \times n$ matrix $\bA$ (also see \cite{DKMH11sigma}).
This is a \emph{matrix version} of the \emph{pole trajectories} in the Belinski--Zakharov approach
\cite{Beli+Zakh78,Beli+Verd01}. A well-known symmetry of the latter extends to the
matrix case: (\ref{tbP_id}) is invariant under $\tbP \mapsto -\epsilon   \tbP^{-1}$.
For the following result, see Lemma~4.1 in \cite{DKMH11sigma}.

\begin{lemma}
\label{lemma:tbP-id}
Any $\theta$-independent, invertible solution of~\eqref{tbP_id}, which commutes with
its derivatives with respect to~$\rho$ and~$\zz$, solves~\eqref{tbP_eqs}.
\end{lemma}

$\bA$ can be taken in Jordan
normal form $\bA = \operatorname{block-diag}(\bA_{n_1},\ldots,\bA_{n_s})$, without restriction of generality,
and a solution of~(\ref{tbP_id}) is then given by
\[
    \tbP = \operatorname{block-diag}(\tbP_{n_1},\ldots,\tbP_{n_s})   ,
\]
where the block $\tbP_{n_i}$
is a solution of (\ref{tbP_id}) with $\bA$ replaced by the Jordan block $\bA_{n_i}$, see the next
examples. Under the assumption that $\tbP$ has geometrically simple spectrum, this is the most general
solution of $\bd \bP = (\d \bP)  \bP$ with $\theta$-independent $\tbP = e^\theta   \bP$~\cite{DKMH11sigma}.

\begin{example}
\label{ex:P_diag}
If $\bA$ is diagonal, i.e., $\bA = \operatorname{diag}(a_1,\ldots,a_n)$ with constants $a_i$,
corresponding solutions of (\ref{tbP_id}) are given by
$\tbP = \operatorname{diag}(\tilde{p}_1,\ldots,\tilde{p}_n)$, where $\tilde{p}_i$ is any of
\begin{gather}
   p_i = \rho^{-1}   ( \zz - a_i + \mathfrak{R}_i )   , \qquad
   \bar{p}_i = \rho^{-1}   ( \zz - a_i - \mathfrak{R}_i )   ,
       \label{p_i}
\end{gather}
with $\mathfrak{R}_i = \sqrt{( \zz - a_i )^2 + \epsilon   \rho^2}$.
Note that $\bar{p}_i = -\epsilon/p_i$.
\end{example}

\begin{remark}
\label{not:p->mu}
For better comparison with the relevant literature,
we will sometimes write
\begin{gather*}
    p_i = \frac{\rho}{\mu_i}   , \qquad
    \bar{p}_i = \frac{\rho}{\bar{\mu}_i}   ,   
\end{gather*}
where
\begin{gather}
    \mu_i = \sqrt{( \zz - a_i )^2 + \epsilon   \rho^2} - (\zz - a_i )   , \qquad
    \bar{\mu}_i = - \sqrt{( \zz - a_i )^2 + \epsilon   \rho^2} - (\zz - a_i)   .  \label{mu,bmu}
\end{gather}
$\mu_i$ and $\bar{\mu}_i$ are often referred to as (indicating the presence of) a \emph{soliton} and
an \emph{anti-soliton}, respectively. Note that $\bar{\mu}_i = - \epsilon   \rho^2/\mu_i$.
If $\epsilon=1$, then $\mu_i$ is non-negative and only vanishes on a~subset of $\{\rho = 0\}$.
\end{remark}

\begin{example}
\label{ex:naCM_non-diag}
For an $r \times r$ Jordan block
\[
       \bA_r = a   \bI_r + \bN_r    , \qquad
       \bN_r = \left( \begin{matrix} 0 & 1 & 0 &   \cdots & 0 \\
                                          0 & 0 & 1 & \ddots         & \vdots \\
                                     \vdots &   & \ddots & \ddots  & \vdots \\
                                   \vdots &   &   &         \ddots & 1  \\
                                        0 & \cdots & \cdots  & \cdots & 0
                    \end{matrix} \right)   ,
\]
(\ref{tbP_id}) has the solutions (see \cite{DKMH11sigma})
\[
    \tbP_r = \rho^{-1}   \left( \zz   \bI_r - \bA_r
     + \sum_{k=0}^{r-1} {1/2 \choose k}
        (\pm \mathfrak{R})^{1-2k} \big[ 2(a-\zz)   \bN_r + \bN_r^2 \big]^k \right)   ,
\]
where $\mathfrak{R} = \sqrt{(\zz-a)^2 + \epsilon  \rho^2}$.
This is an upper-triangular Toeplitz matrix and thus commutes with
its derivatives. In particular, we have
$\tbP_1 = \tilde{p} = \rho^{-1}   [ \zz - a \pm \mathfrak{R} ]$ and
\[
   \tbP_2 = \tilde{p}
        \left( \begin{matrix} 1 & \mp \mathfrak{R}^{-1} \\
           0 & 1 \end{matrix} \right)   , \quad
   \tbP_3 = \tilde{p}
        \left( \begin{matrix} 1 & \mp \mathfrak{R}^{-1} &
           \pm \frac{\epsilon}{2} \rho   \mathfrak{R}^{-3}   \tilde{p}^{-1} \\
           0 & 1 & \mp \mathfrak{R}^{-1} \\
           0 & 0 & 1 \end{matrix} \right)   .
\]
\end{example}

For the solutions $\tbP$ obtained in this way, and thus $\bP$, we have
\[
     \bd \bP = (\d \bP)  \bP = \bP   (\d \bP)   ,
\]
so that they also provide us with solutions $\bQ = e^{-\theta} \tbQ$, with
$\theta$-independent $\tbQ$, of the second of~(\ref{P,Q_eqs}), also see~\cite{DKMH11sigma}.

Recall that on the way to the above results we assumed that $\tbP$ and
$\bI + \epsilon   \tbP^2$ are invertible, and then also
$\tbQ$ and $\bI + \epsilon  \tbQ^2$.
These assumption will also be made throughout in the following.

\begin{remark}
The source matrix $\bA$ for $\tbP$ and the corresponding source matrix~$\bA'$
for~$\tbQ$ can be assumed to be simultaneously in Jordan normal form, without
restriction of generality. Since this is achieved by similarity transformations
with \emph{constant} transformation matrices, the latter can be absorbed by
redef\/initions that restore all the equations obtained from Theorem~\ref{thm:main}.
But in general this will no longer be so if we impose a reduction condition that
relates~$\tbP$ and~$\tbQ$, as in Section~\ref{subsec:sym_reduction} below.
\end{remark}

\subsection[The equations for $U$ and $V$]{The equations for $\bU$ and $\bV$}
\label{subsec:U&V}
Setting $\phi_0 = e^{-\theta} \tphi_0$ with a $\theta$-independent $\tphi_0$
(cf.~(\ref{naCM_phi->tphi})),~(\ref{U,V_eqs})
is autonomous in the variable~$\theta$.
Assuming that $\bU$, $\bV$ are $\theta$-independent, and using the fact that $\tphi_0$ has
to solve the Miura equations~(\ref{naCM_Miura}) together with some~$g_0$,
we obtain the following systems of linear dif\/ferential equations for~$\bU$ and~$\bV$,
\begin{gather}
    \bU_\rho
        =  \big( g_{0,\rho}   g_0^{-1}   \bU - g_{0,z}   g_0^{-1}   \bU \tbP \big)
              \big(\bI + \epsilon   \tbP^2\big)^{-1}   , \nonumber \\
    \bU_z
        =  \big( g_{0,z}   g_0^{-1}   \bU + \epsilon   g_{0,\rho}   g_0^{-1}   \bU \tbP \big)
            \big(\bI + \epsilon  \tbP^2\big)^{-1}   , \label{naCM_U_eqs}
\end{gather}
and
\begin{gather}
    \bV_\rho
        =  \big(\bI + \epsilon   \tbQ^2\big)^{-1}   \big( \tbQ   \bV   g_{0,z}   g_0^{-1}
            - \bV   g_{0,\rho}   g_0^{-1} \big)   , \nonumber \\
    \bV_z
        =  - \big(\bI + \epsilon   \tbQ^2\big)^{-1}   \big( \bV   g_{0,z}   g_0^{-1}
            + \epsilon   \tbQ   \bV   g_{0,\rho}   g_0^{-1} \big)   .
                  \label{naCM_V_eqs}
\end{gather}
These equations
have to be solved for the given \emph{seed solution} $g_0$ of~(\ref{g_eq}).
For \emph{diagonal} $g_0$, this is done in the next example for the
$\bV$-equations. Similar results are easily obtained for the $\bU$-equations.

\begin{example}
\label{ex:Veqs_sols}
Let
\[
  \tbQ = \operatorname{diag}(\tilde{p}_1,\ldots, \tilde{p}_n)   , \qquad
   g_0 = \operatorname{diag}(w_1, \ldots, w_m)   ,
\]
where $\tilde{p}_i$ is either $p_i$ or $\bar{p}_i$ in~(\ref{p_i}),
and $w_\alpha$ is a non-vanishing solution of the scalar
(i.e., $m=1$) version
\begin{gather}
   [\rho   (\ln w)_\rho]_\rho
  = -\epsilon   [\rho  (\ln w)_\zz ]_\zz  \label{naCMscalar}
\end{gather}
of the non-autonomous chiral model (\ref{naCM}).
Writing $\bV = (V_{i\alpha})$, where $i=1,\ldots,n$ and $\alpha=1,\ldots,m$,
(\ref{naCM_V_eqs}) reads
\begin{gather*}
   (\ln V_{i\alpha})_\rho = \frac{1}{1+\epsilon   \tilde{p}_i^2}   \left(
       \tilde{p}_i   (\ln w_\alpha)_z -  (\ln w_\alpha)_\rho \right)   , \\
   (\ln V_{i\alpha})_z = - \frac{1}{1+\epsilon   \tilde{p}_i^2}   \left(
      (\ln w_\alpha)_z + \epsilon   \tilde{p}_i   (\ln w_\alpha)_\rho \right)   .
\end{gather*}
Let us list some simple solutions. If $w_\alpha$ is constant, then also $V_{i\alpha}$.
If $\tilde{p}_i = p_i$, then
\[
 V_{i\alpha} = k_{i\alpha}     \begin{cases}
    (\rho   p_i)^{-1/2} & \mbox{if} \  w_\alpha = \rho, \\
    (\rho   p_i)^{1/2} \big( 1 + \epsilon   p_i^{-1} p_\alpha^{-1}\big) & \mbox{if} \
    w_\alpha = p_\alpha    ,
          \end{cases}
\]
where $p_\alpha$ shall be given by the same expression as some
$p_i$, but with in general dif\/ferent constant, say $a_\alpha'$.
$k_{i\alpha}$ is an arbitrary constant. If $\tilde{p}_i = \bar{p}_i$, then
\[
 V_{i\alpha} =
  k_{i\alpha}   \begin{cases}
     (p_i/\rho)^{1/2} & \mbox{if}  \  w_\alpha = \rho, \\
     (p_i/\rho)^{1/2}  (1 + \epsilon   p_i p_\alpha)^{-1} & \mbox{if}  \
        w_\alpha = p_\alpha    . \end{cases}
\]
More complicated solutions are now obtained by noting the following.
\begin{itemize}\itemsep=0pt
\item If $V_{i\alpha}$ is a solution for $w_\alpha$, then $V_{i\alpha}^{-1}$ is a solution
for $w_\alpha^{-1}$.
\item If $w_\alpha$ is the product of two solutions of (\ref{naCMscalar}),
then $V_{i\alpha}$ is the product of the respective solutions for the factors.
\end{itemize}
\end{example}

\subsection{The Sylvester equation and the solution formula}
\label{subsec:Sylvester}
Recalling that $\bU$, $\bV$ and $\tbQ$ are $\theta$-independent, the formula for
($\theta$-independent) $g$ in (\ref{thm_phi,g}) requires $\bX = e^\theta \tbX$ with
$\theta$-independent $\tbX$.
(\ref{preSylvester}) becomes the $\theta$-independent Sylvester equation
\begin{gather}
    \tbX   \tbP - \tbQ   \tbX = \bV \bU   .  \label{naCM_Sylv}
\end{gather}
If $\mathrm{spec}(\tbP) \cap \mathrm{spec}(\tbQ) = \varnothing$, then
(\ref{naCM_Sylv}) has a unique solution, for any choice of the matrices on the right hand side.
The two matrices $\bP$ and $\bQ$ are then independent, hence Theorem~\ref{thm:main}
implies that\footnote{We have to choose $\tbQ$ invertible and make sure that the
solution $\tbX$ of (\ref{naCM_Sylv}) is invertible. See~\cite{Hearon77,deSouza+Bhatt81}
for conditions that guarantee the latter. }
\begin{gather}
     g = \big(I + \bU \big(\tbQ \tbX\big)^{-1} \bV\big)   g_0   \label{naCM_sol}
\end{gather}
solves the non-autonomous chiral model equation (\ref{naCM}).
Obviously, scaling $\bU$ or $\bV$ with an arbitrary non-zero constant leaves $g$ invariant.
We recall from~\cite{DKMH11sigma} (see Remark~4.4 therein) that
\begin{gather}
     \det g = \frac{\det \tbP}{\det \tbQ}   \det g_0     .   \label{det(g)}
\end{gather}

\begin{example}
\label{ex:Sylv_diag}
Let $\tbP$ be diagonal, as in Example~\ref{ex:P_diag}, and also $\tbQ$, with eigenvalues $\tilde{q}_i$,
given by an expression of the same form as $\tilde{p}_i$.
If they have no eigenvalue in common, i.e., $\{\tilde{p}_i\} \cap \{\tilde{q}_i\} = \varnothing$,
then the unique solution of (\ref{naCM_Sylv}) is given by the Cauchy-like matrix
\begin{gather*}
    \tbX_{ij} = \frac{(\bV \bU)_{ij}}{\tilde{p}_j - \tilde{q}_i}   .   
\end{gather*}
\end{example}

A vast literature exists on solutions of the Sylvester equation (\ref{naCM_Sylv}),
more generally with non-diagonal matrices $\tbP$ and $\tbQ$ (and not necessarily satisfying
the spectrum condition that guarantees a unique solution).

\begin{proposition}[\protect{\cite{Hartwig72,deSouza+Bhatt81,Hu+Cheng06}}]
Let $\mathrm{spec}(\tbP) \cap \mathrm{spec}(\tbQ) = \varnothing$ and
\[
     \mathfrak{P}(\lambda) = \sum_{k=0}^n \mathfrak{P}_k   \lambda^k
\]
be the characteristic polynomial of~$\tbP$. Then the unique solution of the
Sylvester equation \eqref{naCM_Sylv} is given by
\begin{gather}
    \tbX = - \mathfrak{P}(\tbQ)^{-1}   \sum_{k=1}^n \mathfrak{P}_k
             \sum_{i=0}^{k-1} \tbQ^{k-1-i} \bV \bU   \tbP^i   . \label{Sylvester_solution_formula}
\end{gather}
\end{proposition}

\begin{remark}
(\ref{inhomog_ext_lin_sys}) takes the form
\begin{gather}
     \tbX_\rho + \rho^{-1} \tbX + \tbX_z \tbP - \tbQ_z \tbX - \bV_z   \bU  =  0   , \nonumber \\
     \epsilon   \tbX_z - \big(\tbX_\rho - \rho^{-1} \tbX\big)   \tbP + \big(\tbQ_\rho + \rho^{-1} \tbQ\big) \tbX
     + \bV_\rho   \bU  =  0   .  \label{naCM_X_eqs}
\end{gather}
If we drop the spectrum condition for $\tbP$ and $\tbQ$, these equations also have to be
solved. Otherwise they are a consequence of our assumptions (see the proof of Theorem~\ref{thm:main}).
(\ref{naCM_X_eqs}) will only be used in the proof of Proposition~\ref{prop:f_sol}
in Appendix~\ref{app:proofs}.
\end{remark}

\begin{remark}
Using the above results, we also obtain solutions $\phi$ of (\ref{phi_eq}),
given by the expression in (\ref{thm_phi,g}), which in the case under
consideration and via (\ref{naCM_phi->tphi}) takes the form
\[
   \epsilon   \tphi_{zz} + \big(\tphi_{\rho} + \rho^{-1} \tphi\big)_\rho
 = \big[ \tphi_\rho + \rho^{-1} \tphi   ,   \tphi_z \big]
\]
(which corrects a typo in~(4.3) of~\cite{DKMH11sigma}).
\end{remark}

\subsection{A reduction condition}
\label{subsec:sym_reduction}
It is of particular interest (see Section~\ref{sec:Einstein}) to f\/ind a convenient condition
which guarantees that the solution matrix $g$ given by (\ref{naCM_sol}) is \emph{symmetric},
i.e., $g^\intercal = g$, where ${}^\intercal$ means matrix transpose.
The following result is easily verif\/ied by a direct computation.

\begin{lemma}
\label{lemma:sym_red}
Let $\tbP^\intercal = - \epsilon   \tbQ^{-1}$ and $g_0^\intercal = g_0$.
If $\bV$ solves \eqref{naCM_V_eqs}, then $\bU = (\bV   g_0)^\intercal$
solves \eqref{naCM_U_eqs}.
\end{lemma}

\begin{proposition}
\label{prop:sym_red}
Let $\tbP^\intercal = - \epsilon \tbQ^{-1}$, $\mathrm{spec}(\tbP) \cap \mathrm{spec}(\tbQ) = \varnothing$,
$g_0^\intercal = g_0$, and $\bU = (\bV   g_0)^\intercal$.
Then $g$ given by \eqref{naCM_sol} is symmetric.
\end{proposition}

\begin{proof}
Using $g_0^\intercal = g_0$ and $\bU = (\bV   g_0)^\intercal$, the Sylvester equation
(\ref{naCM_Sylv}) and its transpose lead to
\[
      \tbQ \big [ \tbQ \tbX - (\tbQ \tbX)^\intercal \big]
  = \tbQ \big[ \tbX \tbP - \tbP^\intercal \tbX^\intercal \big]
  = \big[ \tbQ \tbX  + \epsilon   \tbX^\intercal \tbP^{-1} \big]   \tbP
  = \big[ \tbQ \tbX - (\tbQ \tbX)^\intercal \big]   \tbP      .
\]
In the last two steps we used $\tbP^\intercal = - \epsilon   \tbQ^{-1}$.
Now the spectrum condition implies $\tbQ \tbX = (\tbQ \tbX)^\intercal$.
Together with $g_0^\intercal = g_0$ and $\bU = (\bV   g_0)^\intercal$, this shows that
$g$ given by~(\ref{naCM_sol}) is symmetric.
\end{proof}

\begin{remark}
Lemma~\ref{lemma:sym_red} and Proposition~\ref{prop:sym_red} also hold with transposition
replaced by any involutory anti-automorphism of the matrix algebra, hence in particular
for Hermitian conjugation.
The \emph{Hermitian reduction} of the non-autonomous chiral model appears in particular
in the context of the (electro-vacuum) Einstein--Maxwell equations in four dimensions
with two commuting Killing vector f\/ields (also see~\cite{DKMH11sigma}).
\end{remark}

In terms of the matrix
\begin{gather}
     \bGamma := -\bQ \tbX = \epsilon \big (\tbP^\intercal\big)^{-1} \tbX   ,  \label{Gamma}
\end{gather}
which is \emph{symmetric} under the assumptions of Proposition~\ref{prop:sym_red}
(see the proof of Proposition~\ref{prop:sym_red}), the Sylvester equation~(\ref{naCM_Sylv})
takes the form of a \emph{Stein equation},
\begin{gather}
    \bGamma + \epsilon   \tbP^\intercal \bGamma \tbP = \bV   g_0   \bV^\intercal   .
        \label{Stein_eq}
\end{gather}
Implementing the assumptions of Proposition~\ref{prop:sym_red} in the solution formula~(\ref{naCM_sol}), we have
\begin{gather}
     g = \big(I - g_0   \bV^\intercal \bGamma^{-1} \bV\big)   g_0   .
          \label{naCM_red_sol}
\end{gather}
If $\tbP$ is diagonal, i.e., $\tbP = \operatorname{diag}(\tilde{p}_1,\ldots,\tilde{p}_n)$, and
if $\tilde{p}_i \tilde{p}_j \neq -\epsilon$ for all $i,j$, then the solution of~(\ref{Stein_eq}) is given
(via Example~\ref{ex:Sylv_diag}) by
\begin{gather}
   \bGamma_{ij} = \epsilon   \frac{(\bV g_0 \bV^\intercal)_{ij}}{\tilde{p}_i \tilde{p}_j + \epsilon}.
           \label{Gamma_Pdiag}
\end{gather}
This is essentially the corresponding matrix (usually denoted by
$\Gamma$) in the Belinski--Zakharov method \cite{Beli+Sakh79,Beli+Verd01}.

\begin{remark}
\label{rem:red_superpos}
 From Remark~\ref{rem:superpos} we deduce the following superposition result.
Let $(\tbP_i, \bV_i)$, $i=1,\ldots,N$, be solutions of (\ref{tbP_eqs}) and (\ref{naCM_V_eqs})
with $\tbQ_i = -\epsilon   (\tbP_i^\intercal)^{-1}$, for the \emph{same}
seed solution~$g_0$. Then $\tbP = \mbox{block-diag}(\tbP_1, \ldots, \tbP_N)$
and $\bV = (\bV_1^\intercal, \ldots, \bV_N^\intercal)^\intercal$ solve
(\ref{tbP_eqs}) and (\ref{naCM_V_eqs}) with $\tbQ = -\epsilon   (\tbP^\intercal)^{-1}$.
If $\mathrm{spec}(\tbP) \cap \mathrm{spec}(-\epsilon (\tbP^\intercal)^{-1})
= \varnothing$, and if~(\ref{Stein_eq}) has an invertible solution $\bGamma$, then~(\ref{naCM_red_sol})
is again a symmetric solution of~(\ref{naCM}).
\end{remark}

\section{Solutions of the vacuum Einstein equations}
\label{sec:Einstein}
In $D$ dimensions, let us consider a space-time metric of the form
\[
    ds^2 = g_{\alpha\beta}   d x^\alpha   d x^\beta + f \big(\epsilon   d \rho^2 + d z^2\big)   ,
\]
where the (real) components $g_{\alpha\beta}$, $\alpha,\beta = 1,\ldots,m$, and the function $f$
only depend on the coordinates $\rho$ and $z$ (and thus not on $x^1,\ldots,x^m$).
The metric then obviously admits $m = D-2$ commuting Killing vector f\/ields\footnote{It is
\emph{not} the most general metric admitting $m = D-2$ commuting Killing vector f\/ields.
See, e.g.,~\cite{Beli+Verd01}.}.
For the next result, see, e.g., \cite{BMG88,Beli+Verd01,Empa+Real08,Tomi+Ishi11}\footnote{There
are also reductions of the Einstein vacuum equations to the non-autonomous chiral model equation
using a \emph{non-Abelian} Lie algebra of Killing vector f\/ields.
See the case with a null (i.e., lightlike, or isotropic) Killing vector f\/ield in four dimensions
treated in \cite{SVV02}.}.
If
\begin{enumerate}\itemsep=0pt
\item[(1)] the matrix $g = (g_{\alpha\beta})$ satisf\/ies $\det g = - \epsilon   \rho^2$,
\item[(2)] $g$ solves the $m \times m$ non-autonomous chiral model equation (\ref{naCM}),
\item[(3)] $f$ is a solution of the compatible system of linear equations
\begin{gather}
   (\ln f)_\rho = - \frac{1}{\rho} + \frac{1}{4 \rho} \operatorname{tr}\big(\cU^2 - \epsilon   \cV^2\big)   ,
   \qquad
   (\ln f)_z = \frac{1}{2 \rho} \operatorname{tr}(\cU \cV)   ,   \label{f_eqs}
\end{gather}
where $\cU := \rho   g_\rho g^{-1}$ and $\cV := \rho   g_z g^{-1}$,
\end{enumerate}
then the metric is Ricci-f\/lat, hence a solution of the vacuum Einstein equations (with
vanishing cosmological constant).

Since $g$ has to be \emph{real} and \emph{symmetric}, corresponding conditions have to be
imposed on the matrix data of the class of solutions obtained in
Section~\ref{sec:naCM}, so that these solutions determine Ricci-f\/lat metrics
(also see~\cite{DKMH11sigma}). Such conditions have been found in Section~\ref{subsec:sym_reduction},
and, accordingly, in all examples of this section we shall set
\begin{gather}
    \tbQ = - \big(\tbP^{-1}\big)^\intercal   , \qquad  \bU = (\bV   g_0)^\intercal    .
                  \label{Einstein_reduction}
\end{gather}
Typically we will regard these equations as def\/ining $\tbQ$ and $\bU$ in terms of
$\tbP$, $\bV$ and $g_0$.
In most of the examples below, $\tbP$ is diagonal.
The (symmetric) solution of the non-autonomous chiral model is
then given by (\ref{naCM_red_sol}) with $\bGamma$ in (\ref{Gamma_Pdiag}).
If $\tbP$ is not diagonal, we have to proceed via the solution
(\ref{Sylvester_solution_formula}) of the Sylvester equation (still assuming that the
spectrum condition holds) and~(\ref{Gamma}).

According to the following remark,
the determinant condition~(1) can always be achieved if~$m$ is odd. There is a~slight restriction if~$m$ is even.

\begin{remark}
\label{rem:g_mod}
By taking the trace of (\ref{naCM}), one f\/inds that
$\det g$ and any power of it is a solution of~(\ref{naCM}) in the scalar case
(also see~\cite{Beli+Zakh78,Beli+Verd01}).
We further note that $\hat{g} = w   g$, with any non-vanishing solution $w$ of the scalar equation,
is again a solution of~(\ref{naCM}).
For a solution~$g$ given by~(\ref{naCM_red_sol}), using~(\ref{det(g)}) and (\ref{Einstein_reduction})
we f\/ind that
\[
    w = \left( \frac{\rho^2}{ (\det \tbP)^2   (-1)^{n+1} \epsilon   \det g_0 } \right)^{1/m}
\]
achieves that $\det \hat{g} = - \epsilon   \rho^2$. Since $w$ has to be real,
for \emph{even} $m$ this requires $(-1)^n (- \epsilon   \det g_0) > 0$.
\end{remark}

In the next subsection, we address~(\ref{f_eqs}). Then we sketch a useful procedure
to construct non-diagonal metrics from diagonal ones in such a way that the diagonal metric is
recovered by setting some parameters to zero.
A collection of relevant examples in four and f\/ive space-time dimensions follows.
The method is indeed of most interest for $D=4$ and $D=5$. The higher the number
of dimensions, the more restrictive is the assumption of $D-2$ commuting Killing vector f\/ields
for the set of solutions of the vacuum Einstein equations.

\subsection[Solutions of the equations for the metric function $f$]{Solutions of the equations for the metric function $\boldsymbol{f}$}

\begin{example}
\label{ex:f_for_diag_g}
 For a \emph{diagonal} solution $g_0$ of (\ref{naCM}),
so that $(g_0)_{\alpha \alpha}$ solves (\ref{naCMscalar}), (\ref{f_eqs}) leads to
\[
     f_0 = \kappa   \rho^{-1} \prod_{\alpha=1}^m \mathfrak{f}_\alpha   ,
\]
where $\kappa$ is an arbitrary constant and $\mathfrak{f}_\alpha$ has to be a solution of
\begin{gather}
 (\ln \mathfrak{f})_\rho = \frac{\rho}{4} \big( (\ln w)_\rho^2 - \epsilon   (\ln w)_z^2 \big)
                    , \qquad
 (\ln \mathfrak{f})_z = \frac{\rho}{2} (\ln w)_\rho   (\ln w)_z   ,
           \label{frakf_eqs}
\end{gather}
with $w$ replaced by $(g_0)_{\alpha \alpha}$. If $w$ is a constant, then also~$\mathfrak{f}$.
Let us write $\mathfrak{f}[w]$ for the solution of the above equations
for a given solution $w$ of~(\ref{naCMscalar}), and let $\propto$ denote equality up to a
non-zero constant factor. In particular, we have
\[
         \mathfrak{f}[\tilde{\mu}_i]
 \propto  \frac{\tilde{\mu}_i}{\sqrt{\tilde{\mu}_i^2 + \epsilon   \rho^2}},
         \qquad \mbox{where} \quad
         \tilde{\mu}_i = \pm \sqrt{(z-a_i)^2 + \epsilon   \rho^2} - (z-a_i)
\]
(cf.~(\ref{mu,bmu})). More generally, we f\/ind
\begin{gather*}
  \mathfrak{f} \left[ \rho^k \frac{ \tilde{\mu}_1 \cdots \tilde{\mu}_r }{ \tilde{\mu}'_1 \cdots \tilde{\mu}'_s} \right]
   \propto   \rho^{k^2/4} \left( \frac{ \tilde{\mu}_1 \cdots \tilde{\mu}_r }{ \tilde{\mu}'_1 \cdots \tilde{\mu}'_s}
                        \right)^{k/2}
       \left(\prod_{i=1}^r \mathfrak{f}[\tilde{\mu}_i] \right)
           \left(\prod_{j=1}^s \mathfrak{f}[\tilde{\mu}'_j] \right)
       \left(\prod_{i < k}  \mathfrak{F}[\tilde{\mu}_i , \tilde{\mu}_k] \right) \\
\hphantom{\mathfrak{f} \left[ \rho^k \frac{ \tilde{\mu}_1 \cdots \tilde{\mu}_r }{ \tilde{\mu}'_1 \cdots \tilde{\mu}'_s} \right]
   \propto }{}
    \times   \left(\prod_{j < l}  \mathfrak{F}[\tilde{\mu}'_j , \tilde{\mu}'_l] \right)
       \prod_i^r \prod_j^s \mathfrak{F}[\tilde{\mu}_i , \tilde{\mu}'_j]^{-1}    ,
\end{gather*}
where $\tilde{\mu}'_i$ is $\tilde{\mu}_i$ with the constant $a_i$ replaced by some
constant $a_i'$, and we introduced the abbreviation
\[
    \mathfrak{F}[\tilde{\mu}_i, \tilde{\mu}_j] = \frac{\tilde{\mu}_i \tilde{\mu}_j}
         {\tilde{\mu}_i \tilde{\mu}_j + \epsilon   \rho^2}
    = \frac{\tilde{\mu}_i - \tilde{\mu}_j}{2   (a_i - a_j)}   ,
\]
where the last equality holds if $a_i \neq a_j$.
The $\tilde{\mu}_i$ (and also the $\tilde{\mu}'_i$) need not be distinct in the above formula.
Using these results in our formula for~$f_0$, we have the solution of~(\ref{f_eqs})
for a large class of diagonal solutions of~(\ref{naCM}).
See Appendix~\ref{app:f} for some details, and also \cite{Beli+Sakh79}.
\end{example}

For the proofs of the following propositions, see Appendix~\ref{app:proofs}.

\begin{proposition}
\label{prop:f_sol}
Let $f_0$ be a solution of \eqref{f_eqs} for a seed $g_0$ $($solution of the Miura
equation$)$.
Let $g$ be a corresponding solution of the non-autonomous chiral model
from the family obtained in Section~{\rm \ref{sec:naCM}} $($i.e., given by \eqref{naCM_sol},
where the ingredients are subject to the respective equations$)$. Then
\[
    f = \kappa   f_0  \rho^{-n} \frac{\det \tbP   \det \tbQ   \det \tbX}{\big[\det\big(\bI + \epsilon   \tbP^2\big)
        \det\big(\bI + \epsilon   \tbQ^2\big)\big]^{1/2}}    ,
\]
with an arbitrary constant $\kappa$, solves \eqref{f_eqs} with the right hand sides evaluated with~$g$.
\end{proposition}

Proposition~\ref{prop:f_sol} does neither assume that~$g_0$ and~$g$ are symmetric, nor the
reduction conditions~(\ref{Einstein_reduction}). The following
corollary now specializes to the case of this reduction.

\begin{corollary}
\label{cor:f_sol}
Let $g_0$ be symmetric, $\tbQ^\intercal = -\epsilon  \tbP^{-1}$,
$\mathrm{spec}(\tbP) \cap \mathrm{spec}(\tbQ) = \varnothing$,
and $\bU = (\bV g_0)^\intercal$.
Then the solution of \eqref{f_eqs} corresponding to $g$ given by \eqref{naCM_red_sol} is
\[
    f = \kappa   f_0   \rho^{-n}  \frac{(\det \tbP)^2}{\det\big(\bI + \epsilon   \tbP^2\big)}    \det \bGamma
          ,
\]
with an arbitrary constant $\kappa$.
\end{corollary}

A convenient formula for the determinant of the solution $\bGamma$ of the
Stein equation seems not to be available in the literature.

If we have to modify a solution $g$ in order to achieve the determinant condition (1),
the following result is of great help.

\begin{proposition}
\label{prop:f_transform}
Let $f$ be a solution of \eqref{f_eqs} for a given solution $g$ from the class obtained in
Section~{\rm \ref{sec:naCM}}, and let $f_{(w)}$ be a solution of~\eqref{f_eqs} for a scalar
solution~$w$ of the non-autonomous chiral model \eqref{naCM}. Then
\begin{gather}
      \hat{f} = (\rho   f_{(w)})^{m-1} \frac{f_{(w   \det g)}}{f_{(\det g)}}   f
                \label{f_tranform}
\end{gather}
solves \eqref{f_eqs} with $\hat{g} = w   g$
$($which also solves \eqref{naCM} according to Remark~{\rm \ref{rem:g_mod})}.
If $\det \hat{g} = -\epsilon   \rho^2$, this reduces to
\begin{gather}
      \hat{f} = \frac{w   f}{(\rho   f_{(w)})^m}   .  \label{f_tranform_special}
\end{gather}
\end{proposition}

Due to these results, the problem of explicitly computing solutions of the vacuum Einstein
equations, based on the results in Section~\ref{sec:naCM} and with a \emph{diagonal} seed,
essentially boils down to that of solving the linear equations (\ref{naCM_V_eqs}) for $\bV$.
For diagonal $\tbQ = -\epsilon  \tbP^{-1}$, solutions of the latter have already been
obtained in Example~\ref{ex:Veqs_sols}.
Although we do not yet have general results in case of a \emph{non}-diagonal~$\tbQ$,
corresponding examples are treated below in Examples~\ref{ex:Q_Jordan} and \ref{ex:TS}.

\subsection{From diagonal to non-diagonal solutions}
\label{subsec:d->nd}

In the following, we impose the reduction conditions (\ref{Einstein_reduction}),
which determine $\tbQ$ in terms of $\tbP$, and $\bU$ in terms of $\bV$ and $g_0$,
and restrict our considerations to the case $\epsilon=1$.

We translate a procedure due to Pomeransky \cite{Pomer06}, which has been formulated and
applied in the Belinski--Zakharov (BZ) approach, to our framework.
We start with a \emph{diagonal} (hence in particular \emph{static}) solution of the above
dimensionally reduced vacuum Einstein equations.
The corresponding solution $\tilde{g}$ of (\ref{naCM}) is thus diagonal.
According to Remark~\ref{rem:scaling} we may still simplify it by multiplication with a
suitable scalar solution of (\ref{naCM}).
Now some (anti-) solitons are removed from it in the following way.
\begin{itemize}\itemsep=0pt
\item Removal of a soliton $p_i = \rho   \mu_i^{-1}$ (i.e., a soliton at $z=a_i$)
from $\tilde{g}_{\alpha\alpha}$.
This means multiplication of $\tilde{g}_{\alpha\alpha}$ by $-p_i^{-2}$. In the BZ language,
the corresponding trivial BZ vector has a~$1$ at the~$\alpha$th position and otherwise zeros.
\item Removal of an anti-soliton $\bar{p}_j = -p_j^{-1} = - \rho^{-1} \mu_j$
(i.e., an anti-soliton at $z=a_j$) from $\tilde{g}_{\beta\beta}$.
This means multiplication of $\tilde{g}_{\beta\beta}$ by $\bar{p}_j^{-2} = -p_j^2$.
\end{itemize}
Afterwards these solitons are reintroduced, with more freedom, in the following way.
\begin{itemize}\itemsep=0pt
\item In order to reintroduce the corresponding soliton, set $\tbP_{kk} = p_i$, for some $k$.
The $k$th row of the matrix $\bV$ that solves (\ref{naCM_V_eqs})
(with $\tbQ^\intercal = - \tbP^{-1}$), or rather
the vector formed by its constant coef\/f\/icients, generalizes the aforementioned trivial BZ vector.
It's  $\alpha$th component has to be non-zero.
\item To reintroduce the corresponding anti-soliton, set $\tbP_{ll} = \bar{p}_j$, for some $l$.
The $l$th row of the matrix $\bV$ generalizes the corresponding trivial BZ vector if its
$\beta$th component is non-zero.
\end{itemize}
After multiplication with a suitable scalar solution of (\ref{naCM}), in order to achieve
the determinant condition, this leads to a non-diagonal (and thus typically stationary)
generalization of the original diagonal metric.
We recover the latter by suitably f\/ixing the parameters introduced in the second step.

In particular, this of\/fers the possibility to reconstruct a known non-diagonal solution
from its diagonal specialization (provided the latter can be achieved)
via the solution-generating technique. Some well-known examples
of four-dimensional space-times are recovered in Section~\ref{subsec:4d} in this way.
The above procedure will be applied in the f\/ive-dimensional case in Section~\ref{subsec:5d}.

\begin{remark}
\label{rem:v=1}
According to an observation at the end of Remark~\ref{rem:scaling}, any transformation
$\bV \mapsto \bW  \bV$ with a constant $n \times n$ matrix $\bW$ that commutes with
$\tbP^\intercal$ (i.e., $[\bW, \tbP^\intercal]=0$) leaves $g$ given by (\ref{naCM_red_sol})
invariant. If $\tbP$ is diagonal, any diagonal $\bW$ commutes with it. We can then use
such a transformation of $\bV$ to scale in each row of $\bV$ one of the non-zero
constant coef\/f\/icients to a f\/ixed value like~$1$. In the above step of reintroduction
of solitons, without restriction of generality we can thus f\/ix the constants appearing
in the respective positions of the rows of~$\bV$.
If $\tbP =\tbP_r$ (see Example~\ref{ex:naCM_non-diag}), then~$\bW$ is an arbitrary constant,
lower-triangular Toeplitz matrix and we can again rescale one non-zero coef\/f\/icient in
each row of the corresponding matrix $\bV$ to a~chosen value (dif\/ferent from zero),
without changing the corresponding solution~$g$.
This feature generalizes to a~$\tbP$ that is block-diagonally composed of matrices~$\tbP_r$.
\end{remark}

\subsection{Solutions of the vacuum Einstein equations in four dimensions}
\label{subsec:4d}
Let $D=4$, hence $m=2$, and $\epsilon=1$. We thus consider stationary and axially
symmetric solutions of the vacuum Einstein equations.
The metric components $g_{\alpha\beta}$, $\alpha,\beta = 1,2$, should then refer to
coordinates $x^1 = t$ (time) and $x^2 = \varphi$ (angle around the symmetry $z$-axis),
$\rho$ is the coordinate distance from the axis. This interpretation in general imposes
additional conditions on the metric. The simple solution
\begin{gather}
     g_0 = \left( \begin{matrix} -1 & 0 \\
                                    0   & \rho^2
                 \end{matrix} \right)
                                       \label{g0_Minkowski}
\end{gather}
of (\ref{naCM}) corresponds to the four-dimensional Minkowski metric
$ds^2 = - dt^2 + \rho^2   d\varphi^2 + d \rho^2 + d z^2$ in cylindrical coordinates.
The solution $f_0$ of (\ref{f_eqs}) with
$g$ given by (\ref{g0_Minkowski}) is simply a constant, which can be set to 1 (since
this can be achieved by a coordinate transformation).

In order to facilitate comparison with the relevant literature, in the following
we will mainly use the $\mu_i$ introduced in (\ref{mu,bmu}),
instead of the $p_i$, see Remark~\ref{not:p->mu}.
Furthermore, we will frequently use the abbreviation (also see \cite{Tomi+Ishi11})
\begin{gather*}
   \cR_{ij} = \rho^2 + \mu_i   \mu_j = 2   (a_i-a_j)   \frac{\mu_i  \mu_j}{\mu_i - \mu_j}   ,         
\end{gather*}
where the last equality holds if $a_i \neq a_j$.

\begin{example}[Kerr-NUT]
\label{ex:Kerr}
In order to recover the Kerr-NUT metric, we start with its static specialization,
the Schwarzschild metric. The latter corresponds to the following solution of~(\ref{naCM}),
\begin{gather*}
     \tilde{g} = \left( \begin{matrix} -\dfrac{\mu_2}{\mu_1} & 0 \\
                                    0   & \rho^2   \dfrac{\mu_1}{\mu_2}
                 \end{matrix} \right)     .
\end{gather*}
Next we remove a soliton at $z=a_1$ from $\tilde{g}_{11}$ and a soliton
at $z=a_2$ from $\tilde{g}_{22}$. The resulting matrix is simplif\/ied by rescaling it
with the inverse of
\[
    w = \rho^{-2}   \mu_1 \mu_2   .
\]
We obtain the seed solution
\[
    g_0 = w^{-1}   \tilde{g}
      \operatorname{diag}\left( -(\rho/\mu_1)^{-2}, -(\rho/\mu_2)^{-2} \right)   ,
\]
which is nothing but (\ref{g0_Minkowski}). Next we reintroduce the two solitons ($n=2$) via
\[
    \tbP = \operatorname{diag}\left( \frac{\rho}{\mu_1} , \frac{\rho}{\mu_2} \right)   .
\]
The respective solution of (\ref{naCM_V_eqs}) (with $\tbQ = - \tbP^{-1}$) is
\[
    \bV = \left( \begin{matrix} v_{11} & v_{12}   \mu_1^{-1} \\
                   v_{21} & v_{22}   \mu_2^{-1}
          \end{matrix} \right)   ,
\]
with constants $v_{i\alpha}$ (see Example~\ref{ex:Veqs_sols}). According to
Remark~\ref{rem:v=1}, we can set $v_{11} = v_{22} = 1$ without restriction of generality.
Let us rename the remaining parameters in $\bV$,
\[
     c_1 = v_{12}   , \qquad c_2 = v_{21}  .
\]
Now we obtain the matrix $\bGamma$ from (\ref{Gamma_Pdiag}). Then $\hat{g} = w   g$,
with $g$ given by (\ref{naCM_red_sol}), satisf\/ies the determinant condition, i.e.,
$\det \hat{g} =-\rho^2$, and has the components
\begin{gather*}
 \hat{g}_{11}  =  \frac{1}{h} \big( \big(c_1^2+ c_2^2\big)   \rho^2   \mu_1 \mu_2   (\mu_1 - \mu_2)^2
                - \mu_1 \mu_2   \big( \big(1+c_1^2 c_2^2\big)   \cR_{12}^2 - 2 c_1 c_2 \cR_{11} \cR_{22}
                  \big) \big)   , \\
 \hat{g}_{12}  =  \frac{(\mu_1 - \mu_2)   \cR_{12}}{h}
          \big( c_1   \big(c_2^2 \mu_2^2 - \rho^2\big)   \cR_{11}
         + c_2 \big(c_1^2 \rho^2 - \mu_1^2\big)   \cR_{22} \big)   , \\
 \hat{g}_{22}  =  \frac{1}{h} \left( - \frac{(\mu_1 - \mu_2)^2}{\mu_1   \mu_2}
         \big( c_1^2   \rho^8 + c_2^2 \mu_1^4 \mu_2^4 \big)
         + \frac{\rho^2}{\mu_1   \mu_2} \big( \big(\mu_1^4 + c_1^2 c_2^2   \mu_2^4\big)   \cR_{12}^2
         - 2 c_1 c_2   \mu_1^2 \mu_2^2   \cR_{11} \cR_{22} \big) \right)   ,
\end{gather*}
where
\[
    h = (\mu_1 - \mu_2)^2 \big(c_1^2 \rho^4 + c_2^2 \mu_1^2 \mu_2^2\big)
       + \big(\mu_1^2 + c_1^2 c_2^2   \mu_2^2\big)   \cR_{12}^2
       - 2  c_1 c_2  \mu_1 \mu_2   \cR_{11} \cR_{22}   .
\]
Furthermore, according to Example~\ref{ex:f_for_diag_g}, we have
\[
     f_{(w)} \propto \frac{\mu_1 \mu_2}{ \sqrt{\cR_{11} \cR_{22}}   \cR_{12} }   .
\]
Corollary~\ref{cor:f_sol} and Proposition~\ref{prop:f_transform} then yield
\[
    \hat{f} =  - \frac{\kappa}{\mu_1 \mu_2   \cR_{11} \cR_{22}}   h    ,
\]
with a constant $\kappa$. Setting $a_2 = - a_1 = \sigma$,
\[
     c_1 = \frac{\mathfrak{m} + \sigma}{\mathfrak{a}+\mathfrak{b}}   , \qquad
     c_2 = \frac{\mathfrak{a}+\mathfrak{b}}{\mathfrak{m} - \sigma}   , \qquad
     \kappa = \mathcal{C}   \frac{(\mathfrak{m}-\sigma)^2}{4   \sigma^2}   , \qquad
     \sigma = \sqrt{\mathfrak{m}^2 - \mathfrak{a}^2 + \mathfrak{b}^2}   ,
\]
with new constants $\mathfrak{a}$, $\mathfrak{b}$, $\mathfrak{m}$ and $\mathcal{C}$,
and passing over from the coordinates~$\rho$ and~$z$ to new coordinates~$r$ and~$\theta$ via
\[
    \rho = \sqrt{(r - \mathfrak{m})^2 - \sigma^2}   \sin \theta   , \qquad
    z = (r - \mathfrak{m})   \cos \theta   ,
\]
we obtain the Kerr-NUT metric in Boyer--Lindquist coordinates,
as given by~(8.48) and~(8.49) in \cite{Beli+Verd01}.
We can compose $N$ Kerr-NUT data blockwise into larger matrices
(the new~$\tbP$ and~$\bV$ are then $2N \times 2N$, respectively $2N \times 2$ matrices),
and obtain again a symmetric solution of the non-autonomous chiral model equation, see
Remark~\ref{rem:red_superpos}. The determinant condition is again achieved by multiplication
with a suitable scalar. In this way we can recover the multi-Kerr-NUT solutions~\cite{Kramer+Neug80}.
\end{example}

Choosing for $\tbP$ the \emph{non-diagonal} solution of~(\ref{tbP_id}) resulting
from an $n \times n$ Jordan block matrix~$\bA_n$ (see Example~\ref{ex:naCM_non-diag}),
we should expect that the resulting family of solutions of the non-autonomous chiral
model can be obtained alternatively via a ``soliton coincidence'' limit (``pole fusion'' in
the Belinski--Zakharov formalism~\cite{Beli+Verd01}) of the family of
solutions obtained with a \emph{diagonal} $\tbP$ (with distinct eigenvalues).
This is conf\/irmed by the following example.

\begin{example}
\label{ex:Q_Jordan}
Again, let $n=2$ and $g_0$ as in~(\ref{g0_Minkowski}). Now we choose
\[
     \tbP = \frac{\rho}{\mu_1} \left( \begin{matrix}
              1 & -\mathfrak{R}^{-1} \\
              0 & 1
              \end{matrix} \right)     ,
\]
where $\mu_1 = \mathfrak{R} - z$ with $\mathfrak{R} = \sqrt{z^2 + \rho^2}$.
This is $\tbP_2$ in Example~\ref{ex:naCM_non-diag} (with $a=0$).
We f\/ind the solution
\[
       \bV = \left( \begin{matrix}
         v_{11} & v_{12}   \mu_1^{-1}  \\
         v_{21} & v_{22}   \mu_1^{-1} - 2 v_{12}   \cR_{11}^{-1}
         \end{matrix} \right)
\]
of (\ref{naCM_V_eqs}), with constants $v_{i\alpha}$. Computing
the solution of the Sylvester equation via~(\ref{Sylvester_solution_formula}),
the symmetric matrix $\bGamma$ is obtained from~(\ref{Gamma}).
We have to rescale the resulting $g$, given by (\ref{naCM_red_sol}), to $\hat{g} = w   g$ with
$w = \rho^{-2}   \mu_1^2$, in order to arrange the determinant condition. It turns out that~$\hat{g}$ only depends on the parameters $v_{i\alpha}$ via
$\alpha = v_{12}/v_{11}$ and $\beta = \det (v_{i\alpha})/v_{11}^2$.
Furthermore, we f\/ind that~$\hat{g}$ also results from the corresponding (Kerr-NUT) solution
in Example~\ref{ex:Kerr} in the limit~$\sigma \to 0$, after setting
$c_1 = \alpha$ and $c_2 = \alpha^{-2} (\alpha - 2 \beta   \sigma)$.
\end{example}

The Kerr-NUT metric consists of two solitons of multiplicity one.
The \emph{Tomimatsu--Sato} metrics~\cite{Tomi+Sato72} are known to generalize it
to two solitons of multiplicity $\delta \geq 1$ (see, e.g.,~\cite{Beli+Verd01})\footnote{The
Tomimatsu--Sato metrics are in fact limiting cases of multiple Kerr-NUT solutions
\cite{Kramer+Neug80,Econ+Tsou89}.}.
We should then expect that, for f\/ixed $\delta$, this solution can be obtained alternatively with
$\tbP$ consisting of two blocks $\tbP_\delta$ as given in Example~\ref{ex:naCM_non-diag}
(corresponding to the $\delta \times \delta$ Jordan block~$\bA_\delta$).
For $\delta=2$ this is conf\/irmed in the next example.

\begin{example}[Tomimatsu--Sato]
\label{ex:TS}
In the static limit, the $\delta=2$ Tomimatsu--Sato metric (also see~\cite{Koda+Hiki03})
reduces to the Zipoy--Voorhees metric. The latter corresponds to the diagonal solution
\[
   \tilde{g} = \operatorname{diag}\left( - \mu_1^{-2} \mu_2^2 , \rho^2 \mu_1^2 \mu_2^{-2} \right)
\]
of (\ref{naCM}).
Now we remove twice a soliton at $z=a_1 = \sigma$ from $\tilde{g}_{11}$, and
also twice a soliton at $z=a_2 = -\sigma$ from $\tilde{g}_{22}$. With a simplifying
scaling we obtain the seed
\[
    g_0 = w^{-1}   \tilde{g}
          \operatorname{diag}\left( \rho^{-4} \mu_1^4 , \rho^{-4} \mu_2^4 \right)   , \qquad
         w = \rho^{-4} \mu_1^2 \mu_2^2   .
\]
The resulting $g_0$ is again (\ref{g0_Minkowski}). Next we reintroduce the two double-solitons
via\footnote{Noting that $p_1/\mathfrak{R}_1 = 2 \rho/\cR_{11}$, this consists of two
blocks of the form $\tbP_2$ in Example~\ref{ex:naCM_non-diag}.}
\[
    \tbP = \left( \begin{matrix}
         \rho   \mu_1^{-1} & - 2   \rho   \cR_{11}^{-1} & 0 & 0 \\
         0 & \rho   \mu_1^{-1} & 0 & 0     \\
         0 & 0 & \rho   \mu_2^{-1} & - 2   \rho   \cR_{22}^{-1} \\
         0 & 0 & 0 & \rho   \mu_2^{-1}
         \end{matrix} \right)   .
\]
The solution of (\ref{naCM_V_eqs}) is then given by
\begin{gather*}
   \bV = \left( \begin{matrix}
         v_{11} & v_{12}   \mu_1^{-1}  \\
         v_{21} & v_{22}   \mu_1^{-1} - 2 v_{12}   \cR_{11}^{-1} \\
         v_{31} & v_{32}   \mu_2^{-1}  \\
         v_{41} & v_{42}   \mu_2^{-1} - 2 v_{32}   \cR_{22}^{-1}
         \end{matrix} \right)  .
\end{gather*}
Again, the matrix $\bGamma$ is obtained via (\ref{Sylvester_solution_formula}) and (\ref{Gamma}).
Next we have to compute $\hat{g} = w   g$, with~$g$ given by (\ref{naCM_red_sol}), and
\[
   \hat{f} = \kappa   \rho^{-6} w   f_{(w)}^{-2}
               \frac{(\det \tbP)^2   \det \bGamma}{\det(\bI + \tbP^2)}, \qquad
   \mbox{where} \quad
   f_{(w)} = \rho^3   \left(\frac{\mu_1^2 \mu_2^2}{\cR_{11} \cR_{12} \cR_{22}}\right)^2   .
\]
According to Remark~\ref{rem:v=1}, without restriction of generality we can set
\[
     v_{11} = v_{21} = v_{32} = v_{42} = 1   .
\]
Choosing
\[
    v_{12} = v_{22} = v_{31} = v_{41}
           = \frac{\mathfrak{q}}{1+\mathfrak{p}}   , \qquad
   \kappa = \sigma^{-2} \left( \frac{1+ \mathfrak{p}}{4  \mathfrak{p}} \right)^4    ,
\]
in terms of prolate spheroidal coordinates $x,y$, given by
\begin{gather}
     \rho = \sigma   \sqrt{(x^2-1)(1-y^2)}   , \qquad
        z = \sigma   x   y   ,     \label{rho,z->x,y}
\end{gather}
and after a coordinate transformation $t \mapsto t - 4 \sigma   \mathfrak{p}^{-1} \mathfrak{q}   \varphi$,
we obtain the $\delta=2$ Tomimatsu--Sato metric
\begin{gather*}
   ds^2 = -\frac{A}{B}   dt^2
     + 8 \sigma \big(1-y^2\big) \frac{\mathfrak{q}   C}{\mathfrak{p}   B}  dt   d\varphi
     + \sigma^2 \big(1-y^2\big) \frac{D}{\mathfrak{p}^2   B}   d\varphi^2\\
\hphantom{ds^2 =}{}
     + \frac{B}{\mathfrak{p}^4 (x^2-y^2)^3} \left( \frac{dx^2}{x^2-1} + \frac{dy^2}{1-y^2} \right)   ,
\end{gather*}
with
\begin{gather*}
   A  =  \big(1-y^2\big)^4 \mathfrak{g}(Z)   \mathfrak{g}(-Z)   , \\
   B  =  \left( \mathfrak{g}(x) + \mathfrak{q}^2 y^4 \right)^2
         + 4 \mathfrak{q}^2 y^2   \left( \mathfrak{p}   x^3 +1
            - ( \mathfrak{p}   x +1)   y^2 \right)^2   , \\
   C  =  \mathfrak{q}^2 ( \mathfrak{p}   x +1) y^4 \big({-}y^2+3\big)
        + \left( -2 \mathfrak{q}^2 \big( \mathfrak{p}   x^3 +1 \big)
                 + ( \mathfrak{p}   x +1)   \mathfrak{g}(x) \right)   y^2
        - \big( 2 \mathfrak{p}   x^3 - \mathfrak{p}   x +1 \big)   \mathfrak{g}(x)   , \\
   D  =   \mathfrak{p}^2 \mathfrak{q}^4 \big(x^2-1\big) y^8
         + 4 \mathfrak{q}^2  ( \mathfrak{p}   x +1) \big(\mathfrak{p}^3   x^3 +3 \mathfrak{p}^2   x^2
             - \mathfrak{p}^3   x + 4 \mathfrak{p}   x - 3 \mathfrak{p}^2 + 4
             \big)   y^6 \\
\hphantom{D=}{}
        -2 \mathfrak{q}^2 \big(3 \mathfrak{p}^4   x^6 + 4 \mathfrak{p}^3   x^5 - 3 \mathfrak{p}^4   x^4
         + 8 \mathfrak{p}^3   x^3 + 37 \mathfrak{p}^2   x^2 - 12 \mathfrak{p}^3   x
         + 48 \mathfrak{p}   x - 13 \mathfrak{p}^2 + 24\big)   y^4  \\
\hphantom{D=}{}
        + 4 \mathfrak{q}^2   \big(\mathfrak{p}^4   x^8 - \mathfrak{p}^4   x^6
            + 4 \mathfrak{p}^3   x^5 - 4 \mathfrak{p}^3   x^3 + 15 \mathfrak{p}^2   x^2
            + 24 \mathfrak{p}   x - 3 \mathfrak{p}^2 +12\big)   y^2 \\
\hphantom{D=}{}
 + \mathfrak{g}(x)   \big( \mathfrak{p}^4 x^6\! + 6 \mathfrak{p}^3 x^5\! - \mathfrak{p}^4 x^4\!
               +16 \mathfrak{p}^2 x^4\! -12 \mathfrak{p}^3 x^3\! + 32 \mathfrak{p}   x^3\!
               +15 \mathfrak{p}^2 x^2\! + 6 \mathfrak{p}^3 x -15 \mathfrak{p}^2\! +16\big)   ,
\end{gather*}
where $\mathfrak{p}^2 + \mathfrak{q}^2 =1$ and
\[
   \mathfrak{g}(x) = \mathfrak{p}^2 x^4 + 2 \mathfrak{p}   x^3 - 2 \mathfrak{p}   x -1
           ,  \qquad
    Z^2 = \frac{x^2-y^2}{1-y^2}   .
\]
These are the expressions (32)--(35) in
\cite{Koda+Hiki03}\footnote{We obtained a factor~$\mathfrak{q}^4$ in the expression
for $D$, whereas there is a factor~$\mathfrak{q}^2$ in~\cite{Koda+Hiki03}. We believe
that the latter is a typo.}.
\end{example}

\begin{remark}
\label{rem:DKM11}
The $m=2$ non-autonomous chiral model is also related to the $D=4$ vacuum Einstein equations
with two commuting Killing vector f\/ields in another way, via the Ernst equation (see, e.g.,~\cite{SKMHH03}).
In this case, one has to impose a dif\/ferent determinant condition: $\det g = 1$.
This allows a \emph{constant} seed $g_0$, and thus \emph{constant} matrices~$\bU$,~$\bV$, in which case the application of Theorem~\ref{thm:main} is
considerably simplif\/ied. The solutions obtained in this way include the multi-Kerr-NUT metrics
(see \cite{DKMH11sigma} for details)\footnote{In \cite{Yazad06,Yazad08}, solutions of the
f\/ive-dimensional Einstein--Maxwell equations are constructed from a pair of solutions
of the $m=2$ non-autonomous chiral model with symmetric $g$ and $\det g = 1$.}.
A similar construction with $m=3$ leads to the (electrically and magnetically) charged
generalizations, the multi-Demianski--Newman metrics~\cite{DKMH11sigma}.
\end{remark}

\subsection{Solutions of the vacuum Einstein equations in f\/ive dimensions}
\label{subsec:5d}
Let $m=3$ and $\epsilon=1$. In this case it seems that no relevant solutions
can be obtained by choosing as the seed a diagonal solution of the non-autonomous chiral model
corresponding to f\/ive-dimensional Minkowski space-time. It then becomes a subtle
problem to choose a suitable seed solution. Here insights about the ``rod structure''
of the putative axis $\rho =0$ are of great help
\cite{Beli+Verd01,Empa+Real02PRD,Harm04,Pomer06,Chen+Teo10}.
Some important solutions that have been obtained or recovered previously in the
Belinski--Zakharov approach will now be presented in our framework.
We use the procedure outlined in Section~\ref{subsec:d->nd}.

\subsubsection[Myers-Perry black holes]{Myers--Perry black holes}

The higher-dimensional generalization of a static black hole is given by the
Schwarzschild--Tangherlini solution~\cite{Tangh63}. The corresponding diagonal
solution of~(\ref{naCM}) is
\[
    \tilde{g} = \operatorname{diag}\left( -\frac{\mu_1}{\mu_2} , \mu_2 , \frac{\rho^2}{\mu_1} \right)  .
\]
The seed $g_0$ is obtained from $\tilde{g}$ as follows:
\begin{itemize}\itemsep=0pt
\item Remove a soliton at $z=a_1$ from $\tilde{g}_{33}$.
\item Remove an anti-soliton at $z=a_2$ from $\tilde{g}_{22}$.
\item Multiply the resulting matrix by $w^{-1}$, with $w = - \mu_1/\mu_2$, to achieve
a simpler form.
\end{itemize}
This results in
\[
     g_0 = \operatorname{diag}\left( 1 , \frac{\rho^2}{\mu_1} , \mu_2 \right)   .
\]
According to Example~\ref{ex:f_for_diag_g}, the associated solution of~(\ref{f_eqs})
is (up to a constant factor)
\[
    f_0 = \frac{\mu_2}{ \sqrt{ \cR_{11} \cR_{22} } }   .
\]
Next we reintroduce the soliton and the anti-soliton via
\[
     \tbP = \operatorname{diag}\left( \frac{\rho}{\mu_1} , -\frac{\mu_2}{\rho}\right )   .
\]
The f\/irst row of the $2 \times 3$ matrix $\bV$ should have a non-zero entry in the
third component, which means we recreate the soliton at $z=a_1$ in the respective
diagonal component of the seed.
The second row of $\bV$ should have a non-zero entry in the second component,
which means we recreate the anti-soliton at $z=a_2$.
According to Example~\ref{ex:Veqs_sols}, the solution of~(\ref{naCM_V_eqs})
(with $\tbQ = - \tbP^{-1}$) is given by
\[
   \bV = \left( \begin{matrix}
         v_{11} & v_{12}   \dfrac{\mu_1}{\cR_{11}} & v_{13}   \dfrac{\cR_{12}}{\mu_1 \mu_2} \vspace{1mm}\\
         v_{21} & v_{22}   \dfrac{\cR_{12}}{\rho^2} & v_{23}  \dfrac{\mu_2}{\cR_{22}}
                \end{matrix} \right)   ,
\]
with constants $v_{i\alpha}$. The corresponding solution of the Stein equation
(\ref{Stein_eq}) is obtained \linebreak from~(\ref{Gamma_Pdiag}). With $g$ given by~(\ref{naCM_red_sol}), the new solution
\[
     \hat{g} = w   g
\]
satisf\/ies the determinant condition: $\det\hat{g} = -\rho^2$.
We f\/ind
\[
     f_{(w)} =  \frac{\cR_{12}}{ \rho   \sqrt{ \cR_{11} \cR_{22} } }
\]
(up to a constant factor). Without restriction of generality, we can set $v_{13} = v_{22} = 1$
(see Remark~\ref{rem:v=1}). With the restrictions and renamings
\[
     v_{12} = v_{23} = 0   , \qquad
     c_1 = v_{11}   , \qquad c_2 = v_{21}   ,
\]
use of Corollary~\ref{cor:f_sol} and Proposition~\ref{prop:f_transform} yields
\[
    \hat{f} =  \kappa   \mu_2   \frac{c_1^2 c_2^2   \mu_1^3   \mu_2
       - (\mu_1-\mu_2)^2   \left( \mu_1   \big(c_1^2 \mu_1 \mu_2 + c_2^2 \rho^2\big) + \cR_{12}^2 \right) }
         { (\mu_1-\mu_2)^2   \cR_{11} \cR_{12} \cR_{22} }   ,
\]
with a constant $\kappa$.
The corresponding metric is given by
\[
    ds^2 = \hat{g}_{11}   dt^2 + \hat{g}_{22}   d\varphi^2 + \hat{g}_{33}   d\psi^2
             + 2   \hat{g}_{12}   dt   d\varphi + 2   \hat{g}_{13}   dt   d\psi
             + 2   \hat{g}_{23}   d\varphi   d\psi + \hat{f}   \big(d\rho^2 + dz^2\big)   .
\]
Setting
\[
    a_1 = - \sigma    , \qquad a_2 = \sigma  ,
\]
with a constant $\sigma$, in terms of prolate spheroidal coordinates $x$, $y$, given by (\ref{rho,z->x,y}),
we have
\begin{gather*}
   \hat{g}_{11}  =  - \left( 4 \sigma   x + \big( \mathfrak{a}_1^2 - \mathfrak{a}_2^2 \big)   y
                    - \rho_0^2 \right)   \omega^{-1}     , \\
   \hat{g}_{12}  =  - \mathfrak{a}_1   \rho_0^2   (1-y)   \omega^{-1}    , \\
   \hat{g}_{13}  =  - \mathfrak{a}_2   \rho_0^2   (1+y)   \omega^{-1}    , \\
   \hat{g}_{22}  =  \frac{1-y}{4} \left( 4 \sigma   x + \mathfrak{a}_1^2 - \mathfrak{a}_2^2
                    + \rho_0^2 + 2 \mathfrak{a}_1^2 \rho_0^2   (1-y)   \omega^{-1} \right)   , \\
   \hat{g}_{23}  =  \frac{1}{2} \mathfrak{a}_1 \mathfrak{a}_2   \rho_0^2   \big(1 - y^2\big)
                      \omega^{-1}   , \\
   \hat{g}_{33}  = \frac{1+y}{4} \left( 4 \sigma   x - \mathfrak{a}_1^2 + \mathfrak{a}_2^2
                    + \rho_0^2 + 2 \mathfrak{a}_2^2 \rho_0^2   (1+y)   \omega^{-1} \right)   , \\
        \hat{f}  =  \frac{\omega}{ 8 }    ,  \qquad
        \omega = 4 \sigma   x + (\mathfrak{a}_1^2 - \mathfrak{a}_2^2)   y + \rho_0^2  .
\end{gather*}
Performing a linear transformation of the coordinates $t$, $\varphi$, $\psi$,
\begin{gather*}
 (t,\varphi,\psi) \mapsto
    \left( t-\mathfrak{a}_1   \varphi -\mathfrak{a}_2   \psi   ,
      - \frac{\mathfrak{a}_1}{c_2}   \varphi
              + \frac{\mathfrak{a}_2   c_2}{4   \sigma}   \psi   ,
      \frac{\mathfrak{a}_1   c_1}{4   \sigma}  \varphi - \frac{\mathfrak{a}_2}{c_1}   \psi  \right)   ,
\end{gather*}
and setting
\begin{alignat*}{3}
  &  \sigma = \frac{1}{4} \sqrt{(\rho_0^2 -\mathfrak{a}_1^2 -\mathfrak{a}_2^2)^2
                   - 4 \mathfrak{a}_1^2 \mathfrak{a}_2^2}   , \qquad &&
      \kappa = \frac{(\mathfrak{a}_1^2 -\mathfrak{a}_2^2)^2 - (4 \sigma + \rho_0^2)^2}
                { 16   \sigma   \rho_0^2 }   , & \\
 &   c_1^2 = \left| \frac{ 4 \sigma   (\rho_0^2 - \mathfrak{a}_1^2 + \mathfrak{a}_2^2 - 4 \sigma)}
                   {\rho_0^2 + \mathfrak{a}_1^2 - \mathfrak{a}_2^2 + 4 \sigma)} \right|   , \qquad &&
      c_2^2 = \left| \frac{4 \sigma   ( \rho_0^2 + \mathfrak{a}_1^2 - \mathfrak{a}_2^2 - 4 \sigma)}
                   {\rho_0^2 - \mathfrak{a}_1^2 + \mathfrak{a}_2^2 + 4 \sigma} \right|   , &
\end{alignat*}
we recover the analogue of the Kerr metric, i.e., the Myers--Perry solution
\cite{Myers+Perry86,Harm04,Myers12} of the f\/ive-dimensional vacuum Einstein equations,
precisely in the form of equation~(19) in~\cite{Pomer06}.
Switching on the constants~$v_{12}$ and~$v_{23}$ (which we set to zero above),
leads to a more general class of space-times.

\subsubsection{Black saturn}
A black saturn \cite{Elva+Figu07} (also see \cite{Yazad07,CES10,Szyb11}) is a black hole
surrounded by a black ring\footnote{Black rings are similar to black holes,
but with horizon topology $S^1 \times S^{D-3}$. They only appear in $D >4$ dimensions.
See \cite{Empa+Real02PRL,Hong+Teo03,TUS04,Gaun+Guto05,Figueras05,Empa+Real06,Pomer+Senkov06,Mish+Iguc06,Iguchi+Mishima06,
TMY06,Tomi+Noza06,Frolov07,Elva+Rodr08,Izumi08,Holl+Yaza08,Hoski09,Evsl+Kris09,Chru+Cort10,CHT11}.}.
In the following, we show how the black saturn solution, originally obtained in~\cite{Elva+Figu07},
can be recovered in our approach. Let us start with the same static solution of the
non-autonomous chiral model as in~\cite{Elva+Figu07},
\[
   \tilde{g} = \operatorname{diag}\left( -\frac{\mu_1 \mu_3}{\mu_2 \mu_4} , \frac{\rho^2 \mu_4}{\mu_3 \mu_5} ,
          \frac{\mu_2 \mu_5}{\mu_1}\right)   .
\]
This is motivated by a rod structure analysis (also see
\cite{Empa+Real02PRD,Harm04,Chen+Teo10}). A suitable seed solution is then constructed as
follows.
\begin{itemize}\itemsep=0pt
\item Remove anti-solitons at $z=a_1$ and at $z=a_3$ from $\tilde{g}_{11}$.
\item Remove a soliton at $z=a_2$ from $\tilde{g}_{11}$.
\item Multiply the resulting matrix by $w^{-1}$ with $w = \frac{\rho^2 \mu_2}{\mu_1 \mu_3}$
to simplify its form.
\end{itemize}
This results in
\[
   g_0 = w^{-1}   \tilde{g}   \operatorname{diag}\left( -\frac{\rho^2 \mu_2^2}{\mu_1^2 \mu_3^2} , 1 , 1 \right)
       = \operatorname{diag}\left( \frac{1}{\mu_4} , \frac{\mu_1 \mu_4}{\mu_2 \mu_5} , \frac{\mu_3 \mu_5}{\rho^2} \right)
      .
\]
The corresponding solution of (\ref{f_eqs}) is obtained via Example~\ref{ex:f_for_diag_g},
\[
  f_0
     = k^2   \mu_3   \mu_4   \mu_5   \frac{\cR_{12} \cR_{15} \cR_{24} \cR_{45} }
       { \sqrt{ \cR_{11} \cR_{22} \cR_{33} }   \cR_{14} \cR_{25} \cR_{35} \cR_{44} \cR_{55} }   ,
\]
with a constant $k$. We have $n=3$ and choose $\tbP$ diagonal with
\[
     \tbP_{11} = -\mu_1/\rho   , \qquad
     \tbP_{22} = \rho/\mu_2   , \qquad
     \tbP_{33} = -\mu_3/\rho   .
\]
The solution of the equations for $\bV$ is then given by
\[
    \bV = \left( \begin{matrix} v_{11}
          \dfrac{\cR_{14}}{\mu_1}
              & v_{12}   \dfrac{ \cR_{12} \cR_{15} }{ \cR_{11} \cR_{14} }
                & v_{13}   \dfrac{ \rho^2 \mu_1 }{ \cR_{13} \cR_{15} } \vspace{1mm}\\
          v_{21}   \dfrac{ \mu_2 \mu_4 }{ \cR_{24} }
              & v_{22}   \dfrac{ \mu_2 \mu_5   \cR_{12} \cR_{24} }
                  { \mu_1 \mu_4   \cR_{22} \cR_{25} }
                & v_{23}   \dfrac{\mu_1   \cR_{23} \cR_{25} }{ \mu_2^2 \mu_3 \mu_5 }  \vspace{1mm}\\
          v_{31}   \dfrac{ \cR_{34} }{ \mu_3 }
              & v_{32}   \dfrac{ \cR_{23} \cR_{35} }{ \cR_{13} \cR_{34} }
                & v_{33}   \dfrac{\rho^2 \mu_3}{ \cR_{33} \cR_{35} }
                 \end{matrix} \right)   ,
\]
with constants $v_{i\alpha}$. Without restriction of generality, we can set
\[
   v_{11} = v_{21}= v_{31} = 1   .
\]
The case considered in Section~2.2 of \cite{Elva+Figu07}
should then correspond to the subclass of solutions given by
\[
     v_{12} = v_{22} = v_{33} = 0    .
\]
The authors of \cite{Elva+Figu07} then only elaborate the special case
\[
     v_{32} = 0
\]
further. In this case, however, the removal of the anti-soliton at $z=a_3$ from the original
static metric and the subsequent reintroduction via~$\tbP_{33}$ is actually redundant.
This means that the black saturn space-time can already be obtained from $n=2$ data.
The solution $\bGamma$ of the Stein equation~(\ref{Stein_eq}) is given by~(\ref{Gamma_Pdiag}). With $g$ given by~(\ref{naCM_red_sol}), we set
$\hat{g} = w   g$ to satisfy the determinant condition.
Then $\hat{g}$ reduces to $\tilde{g}$ if $v_{13} = v_{23} =0$, as expected.
We f\/ind
\[
     f_{(w)}
  = \frac{ \cR_{12} \cR_{23} }{ \sqrt{ \cR_{11} \cR_{22} \cR_{33} }    \cR_{13} }
\]
(up to a constant factor) and, using Corollary~\ref{cor:f_sol} and Proposition~\ref{prop:f_transform},
\[
     \hat{f} = -\kappa   \rho^{-6} w   f_0   f_{(w)}^{-3}
               \frac{(\det \tbP)^2   \det \bGamma}{\det\big(\bI + \tbP^2\big)}   ,
\]
which turns out to be a lengthy expression. Setting
\[
    v_{13} = 2 c_1   (a_1-a_5)   , \qquad
    v_{23} = c_2   [ 4 (a_2-a_3) (a_2-a_4) ]^{-1}   , \qquad
    \kappa = 4 \frac{(a_1-a_2)^2 (a_2-a_3)^2}{(a_1-a_3)^2}   ,
\]
with constants $c_i$, we obtain the metric
\[
    ds^2 = - \frac{H_2}{H_1} \left( dt + \frac{\omega}{H_2}   d \psi \right)^2
           + H_1 \left( k^2 \mathcal{P}  ( d \rho^2 + dz^2) + \frac{G_1}{H_1}   d\varphi^2
           + \frac{G_2}{H_2}   d \psi^2 \right)   ,
\]
where
\begin{gather*}
   G_1  =  \frac{\rho^2 \mu_4}{\mu_3 \mu_5}   , \qquad G_2 = \frac{\mu_3 \mu_5}{\mu_4}   , \qquad
           \mathcal{P} = \cR_{15} \cR_{34}^2 \cR_{45}   , \\
   H_1  =  F^{-1} \left( M_0 + c_1^2 M_1 + c_2^2 M_2 + c_1 c_2   M_3 + c_1^2 c_2^2   M_4 \right)   , \\
   H_2  =  \frac{\mu_3}{\mu_4   F} \left( \frac{\mu_1}{\mu_2}  M_0
           - c_1^2 \frac{\rho^2}{\mu_1 \mu_2}   M_1
           - c_2^2 \frac{\mu_1 \mu_2}{\rho^2}   M_2
           + c_1 c_2 M_3 + c_1^2 c_2^2  \frac{\mu_2}{\mu_1}   M_4 \right)   , \\
   F  =  \mu_1 \mu_5 (\mu_1-\mu_3)^2 (\mu_2-\mu_4)^2 \left(\prod_{i=1}^5 \cR_{ii} \right)
         \cR_{13} \cR_{14} \cR_{23} \cR_{24} \cR_{25} \cR_{35}   ,
\end{gather*}
with
\begin{gather*}
   M_0  =  \mu_2 \mu_5^2   (\mu_1-\mu_3)^2 (\mu_2-\mu_4)^2   \cR_{12}^2
             \cR_{14}^2 \cR_{23}^2   , \\
   M_1  =  \rho^2 \mu_1^2 \mu_2 \mu_3 \mu_4 \mu_5   (\mu_1-\mu_2)^2 (\mu_1-\mu_5)^2 (\mu_2-\mu_4)^2
             \cR_{23}^2   , \\
   M_2 = \rho^2 \mu_1^2 \mu_2^{-1} \mu_3 \mu_4 \mu_5   (\mu_1-\mu_2)^2 (\mu_1-\mu_3)^2
             \cR_{14}^2 \cR_{25}^2    , \\
   M_3 = 2 \mu_1^2 \mu_3 \mu_4 \mu_5   (\mu_1-\mu_3) (\mu_1-\mu_5) (\mu_2-\mu_4)
           \cR_{11} \cR_{22} \cR_{14} \cR_{23} \cR_{25}   , \\
   M_4 = \mu_1^4 \mu_2^{-1} \mu_3^2 \mu_4^2   (\mu_1-\mu_5)^2
           \cR_{12}^2 \cR_{25}^2   , \\
   \omega = \frac{2}{F   \sqrt{G_1}} \left( c_1 \mathfrak{R}_1 \sqrt{M_0 M_1}
          - c_2 \mathfrak{R}_2 \sqrt{M_0 M_2} + c_1^2 c_2 \mathfrak{R}_2 \sqrt{M_1 M_4}
          - c_1 c_2^2 \mathfrak{R}_1 \sqrt{M_2 M_4} \right)   ,
\end{gather*}
and $\mathfrak{R}_i = \sqrt{(z-a_i)^2+\rho^2}$. This is the black saturn
metric\footnote{Additional conditions have to be imposed on the remaining
parameters in order to achieve asymptotic f\/latness and absence of naked and conical
singularities, see \cite{Elva+Figu07,CES10}. }
as given in \cite{Elva+Figu07,Tomi+Ishi11}, with some obvious changes in notation,
but some deviations in the factors $\mu_i$ in the expressions for~$M_2$,~$M_3$ and~$M_4$.

\subsubsection[Double Myers-Perry black hole solution]{Double Myers--Perry black hole solution}
\label{subsec:dMP}
In order to recover the \emph{double} Myers--Perry black hole solution obtained in~\cite{HRZC08},
we start with the matrix that determines a \emph{static} two-black hole solution (cf.~(3.1) in~\cite{Tan+Teo03}),
\[
     \tilde{g} = \operatorname{diag}\left( -\frac{\mu_1 \mu_4}{\mu_2 \mu_5} ,
         \frac{\rho^2 \mu_3}{\mu_1 \mu_4} , \frac{\mu_2 \mu_5}{\mu_3} \right)   .
\]
Removal of two solitons $\mu_2$, $\mu_5$ and two anti-solitons $\mu_1$, $\mu_4$ from $\tilde{g}_{11}$,
and simplif\/ication with a~suitable factor, leads to
\[
   g_0 = w^{-1}   \tilde{g}
          \operatorname{diag}\left( -\left(\frac{\mu_1^2 \mu_4^2}{\mu_2^2 \mu_5^2}\right)^{-1} , 1 , 1 \right)
       = \operatorname{diag}\left( -1 , \frac{\rho^2 \mu_3}{\mu_2 \mu_5} , \frac{\mu_1 \mu_4}{\mu_3} \right)
         , \qquad w = \frac{\mu_2 \mu_5}{\mu_1 \mu_4}   .
\]
According to Example~\ref{ex:f_for_diag_g}, the corresponding solution of (\ref{f_eqs}) is
given by
\[
    f_0 = \frac{\mu_1 \mu_4}{\mu_3}   \left( \prod_{i=1 \atop i\neq 3}^5 \sqrt{ \cR_{ii} } \right)^{-1}
          \frac{ \cR_{13} \cR_{23} \cR_{34} \cR_{35} }
          { \cR_{14} \cR_{25} \cR_{33} }   .
\]
We have $n=4$ and reintroduce the removed solitons and anti-solitons via
\[
   \tbP = \operatorname{diag}\left( - \rho^{-1} \mu_1 , - \rho^{-1} \mu_4 , \rho   \mu_2^{-1} ,
                         \rho   \mu_5^{-1} \right)   .
\]
The solution of (\ref{naCM_V_eqs}) is given by
\[
   \bV = \left( \begin{matrix}
          v_{11}
            & v_{12}   \dfrac{ \cR_{12} \cR_{15} }{ \rho^2 \cR_{13}}
               & v_{13}   \dfrac{ \mu_1   \cR_{13} }{ \cR_{11} \cR_{14} }    \vspace{1mm}\\
          v_{21}
            & v_{22}   \dfrac{ \cR_{24} \cR_{45} }{ \rho^2 \cR_{34} }
                & v_{23}   \dfrac{ \mu_4   \cR_{34} }{ \cR_{14} \cR_{44} }   \vspace{1mm}\\
          v_{31}
            & v_{32}   \dfrac{ \mu_2 \mu_5   \cR_{23} }{ \mu_3   \cR_{22} \cR_{25} }
                & v_{33}   \dfrac{ \mu_3   \cR_{12} \cR_{24} }{ \mu_1 \mu_2 \mu_4   \cR_{23} } \vspace{1mm} \\
          v_{41}
            & v_{42}   \dfrac{ \mu_2 \mu_5   \cR_{35} }{ \mu_3   \cR_{25} \cR_{55} }
                & v_{43}   \dfrac{ \mu_3   \cR_{15} \cR_{45} }{ \mu_1 \mu_4 \mu_5   \cR_{35} }
                 \end{matrix} \right)  .
\]
In order to recover the double Myers--Perry black hole solution presented in \cite{HRZC08},
we reduce the set of solutions obtained in this way by restricting $\bV$ to
\[
   \bV = \left( \begin{matrix}
          1
            & v_{12}   \dfrac{ \cR_{12} \cR_{15} }{ \rho^2 \cR_{13}}
               & 0   \vspace{1mm}\\
          1
            & v_{22}   \dfrac{ \cR_{24} \cR_{45} }{ \rho^2 \cR_{34} }
                & 0  \vspace{1mm}\\
          1 & 0 & 0  \\
          1 & 0 & 0
                 \end{matrix} \right)   .
\]
We immediately notice that this means in particular ``trivializing'' the solitons at $z=a_2$ and
$z=a_5$. Hence, the solution obtained with this special choice of~$\bV$ can already
be obtained from $n=2$ data.
We should expect, however, a $4$-soliton transformation to be necessary in order to
generate a (suf\/f\/iciently general) double black hole solution, which suggests to
explore the above more general solution. This will not be done here, and we
return to the special case with the above restricted $\bV$.
Again, we obtain $\bGamma$ from~(\ref{Gamma_Pdiag}). Let
$\hat{g}$ be the resulting solution~(\ref{naCM_red_sol}), multiplied by $w$
to achieve the determinant condition. From Example~\ref{ex:f_for_diag_g}, we obtain
\[
    f_{(w)}
  = \rho^{-1} \left( \prod_{i=1 \atop i \neq 3}^5 \sqrt{ \cR_{ii} } \right)^{-1}
        \frac{ \cR_{12} \cR_{15} \cR_{24} \cR_{45} }{ \cR_{14} \cR_{25} }   ,
\]
and then $\hat{f}$ via Corollary~\ref{cor:f_sol} and Proposition~\ref{prop:f_transform}.
Setting
\begin{gather*}
     v_{12} = \frac{b   (a_1 - a_3)}{2 (a_1 - a_2)(a_1 - a_5)}   , \qquad
      v_{22} = \frac{c   (a_3 - a_4)}{2 (a_2 - a_4) (a_4 - a_5) }   , \\
     \kappa = \left(4 \frac{(a_1 - a_2)(a_1 - a_5)(a_2 - a_4)(a_4 - a_5) }
             { (a_1 - a_4) (a_2 - a_5) }\right)^2   ,
\end{gather*}
with constants $b$, $c$, this results in the metric
\[
    ds^2 = - \frac{H_2}{H_1} \left( dt + \frac{\omega}{H_2}   d \varphi \right)^2
           + \frac{\rho^2 \mu_3   H_1}{\mu_2 \mu_5   H_2}   d \varphi^2 + \frac{\mu_2 \mu_5}{\mu_3}   d \psi^2
           + k   \frac{H_1}{F}   \big( d \rho^2 + dz^2\big)   ,
\]
where
\begin{gather*}
   H_1  =  M_0 + b^2 M_1 + c^2 M_2 + b c   M_3 + b^2 c^2   M_4   , \\
   H_2  =  \frac{\rho^2}{\mu_2 \mu_5} \left( \frac{\mu_1 \mu_4}{\rho^2}   M_0
           - b^2 \frac{\mu_4}{\mu_1} M_1 - c^2 \frac{\mu_1}{\mu_4} M_2
           - b c   M_3 + b^2 c^2 \frac{\rho^2}{\mu_1 \mu_4}   M_4 \right)    , \\
   F  =  \mu_3^2   (\mu_1-\mu_4)^2 \left( \prod_{i=1}^5 \cR_{ii} \right)
         \frac{ \cR_{12} \cR_{14}^2 \cR_{15} \cR_{24} \cR_{25}^2 \cR_{45} }
         { \cR_{13} \cR_{23} \cR_{34} \cR_{35} }    , \\
   \omega  =  - 2   \left( \frac{\mu_3}{\mu_2 \mu_5} \right)^{1/2} \left( b   \mathfrak{R}_1 \sqrt{M_0 M_1}
              + c   \mathfrak{R}_4 \sqrt{M_0 M_2} - b^2 c   \mathfrak{R}_4 \sqrt{M_1 M_4}
              - b c^2 \mathfrak{R}_1 \sqrt{M_2 M_4} \right)     ,
\end{gather*}
with
\begin{gather*}
   M_0  =  \mu_2 \mu_3^2 \mu_5   (\mu_1-\mu_4)^2   \cR_{12}^2 \cR_{15}^2 \cR_{24}^2 \cR_{45}^2   , \\
   M_1  =  \mu_1^2 \mu_2^2 \mu_3 \mu_5^2   (\mu_1-\mu_3)^2   \cR_{14}^2 \cR_{24}^2 \cR_{45}^2   , \\
   M_2  =  \mu_2^2 \mu_3 \mu_4^2 \mu_5^2   (\mu_3-\mu_4)^2   \cR_{12}^2 \cR_{14}^2 \cR_{15}^2   , \\
   M_3  =  2 \mu_1 \mu_2^2 \mu_3 \mu_4 \mu_5^2   (\mu_1-\mu_3) (\mu_3-\mu_4)
           \cR_{11} \cR_{12} \cR_{15} \cR_{24} \cR_{44} \cR_{45}   , \\
   M_4  =  \rho^4 \mu_1^2 \mu_2^3 \mu_4^2 \mu_5^3   (\mu_1-\mu_3)^2 (\mu_1-\mu_4)^2 (\mu_3-\mu_4)^2   .
\end{gather*}
With obvious changes in notation, this is the metric obtained in \cite{HRZC08}.

\begin{remark}
It is plausible that one can start with the diagonal solution of the non-autonomous chiral
model corresponding to a static triple black hole space-time (see~(4.1) in~\cite{Tan+Teo03}) and
construct a space-time with three Myers--Perry black holes. This procedure should continue
to produce solutions with an arbitrary number of rotating black holes.
\end{remark}

\subsubsection{Bicycling black rings}\label{subsec:bbr}

Let us start with the solution
\[
     \tilde{g} = \operatorname{diag}\left( -\frac{\mu_1 \mu_5}{\mu_3 \mu_7} ,
          \frac{\rho^2 \mu_3 \mu_7}{\mu_2 \mu_4 \mu_6} , \frac{\mu_2 \mu_4 \mu_6}{\mu_1 \mu_5} \right)
\]
of (\ref{naCM}), which corresponds to a static metric. Removal of a soliton at $z=a_7$
and an anti-soliton at $z=a_1$ from $\tilde{g}_{11}$, and a rescaling,
leads to the seed metric
\[
    g_0 = w^{-1}   \tilde{g}   \operatorname{diag}\left( \frac{\mu_7^2}{\mu_1^2},1,1 \right)
        = \operatorname{diag}\left( \frac{\mu_5}{\mu_3}, - \frac{\rho^2 \mu_1 \mu_3}{\mu_2 \mu_4 \mu_6} ,
          - \frac{\mu_2 \mu_4 \mu_6}{\mu_5 \mu_7} \right)   , \qquad w = -\frac{\mu_7}{\mu_1}   ,
\]
with the following solution of (\ref{f_eqs}) (up to a constant factor),
\[
    f_0 = \frac{ \mu_2 \mu_4 \mu_6 }{ \mu_5 \mu_7 }   \frac{ \cR_{12} \cR_{14} \cR_{16}
          \cR_{23} \cR_{25} \cR_{27} \cR_{34} \cR_{35} \cR_{36} \cR_{45} \cR_{47} \cR_{56} \cR_{67} }
          { \sqrt{\cR_{11}} \cR_{13} \cR_{22} \cR_{24}^2 \cR_{26}^2 \cR_{33} \cR_{44}
            \cR_{46}^2 \cR_{55} \cR_{57} \cR_{66} \sqrt{\cR_{77}} }   .
\]
We have $n=2$ and shall set
\[
    \tbP = \operatorname{diag}\big({-} \rho^{-1} \mu_1 , \rho   \mu_7^{-1} \big)   .
\]
The solution of (\ref{naCM_V_eqs}) is then given by (see Example~\ref{ex:Veqs_sols})
\[
 \bV = \left( \begin{matrix}
       v_{11}   \dfrac{\cR_{13}}{ \cR_{15} }
         & v_{12}   \dfrac{ \cR_{12} \cR_{14}\cR_{16} }{ \mu_1 \cR_{11} \cR_{13} }
           & v_{13}   \dfrac{ \mu_1 \cR_{15} \cR_{17} }{ \cR_{12} \cR_{14} \cR_{16} } \vspace{1mm}\\
       v_{21}   \dfrac{ \mu_3 \cR_{57} }{ \mu_5 \cR_{37} }
         & v_{22}   \dfrac{\mu_2 \mu_4 \mu_6 \cR_{17} \cR_{37} }{ \mu_1 \mu_3 \cR_{27} \cR_{47} \cR_{67} }
           & v_{23}   \dfrac{\mu_5 \cR_{27} \cR_{47} \cR_{67}}{ \mu_2 \mu_4 \mu_6 \cR_{57} \cR_{77} }
                 \end{matrix} \right)   .
\]
Without restriction of generality, we can set $v_{11}=v_{21}=1$.
But we do restrict the class of solutions by setting
\[
   v_{12} = v_{23}=0   .
\]
Again, the solution $\bGamma$ of the Stein equation is obtained from (\ref{Gamma_Pdiag}).
The resulting solution~(\ref{naCM_red_sol}) of the non-autonomous chiral model
has to be modif\/ied to $\hat{g} = w   g$, with $w$ as given above,
in order to achieve the determinant condition. We have (disregarding a constant factor)
\[
     f_{(w)} = \rho^{-1}   \frac{ \cR_{17} }{ \sqrt{\cR_{11}   \cR_{77}} }    ,
\]
and, from Corollary~\ref{cor:f_sol} and Proposition~\ref{prop:f_transform},
\[
    \hat{f} = \kappa   \rho^{-5} w   f_0   f_{(w)}^{-3}
              \frac{(\det \tbP)^2   \det \bGamma}{\det\big(\bI+\tbP^2\big)}   ,
\]
which results in a lengthy expression. Setting
\[
   v_{13} = c_1   , \qquad
   v_{22} = b_2   \frac{ (a_7-a_4)(a_7-a_5)(a_7-a_6)}{(a_7-a_1)(a_7-a_3)^2}   , \qquad
   \kappa = 4   (a_7-a_1)^2   ,
\]
we obtain the metric\footnote{Here we used
$a_i - a_j = (\mu_i - \mu_j)(\rho^2+\mu_i \mu_j)/(2 \mu_i \mu_j)$
to eliminate $a_i-a_j$. }
\begin{gather*}
    ds^2 = - \frac{H_2}{H_1} \left( dt - \frac{\omega_1}{H_2}   d \varphi
               - \frac{\omega_2}{H_2}   d \psi \right)^2\\
\hphantom{ds^2 =}{}
           + \frac{1}{H_2} \left( G_1   d \varphi^2 + G_2   d \psi^2 - 2 J   d\varphi   d \psi \right)
           + \cP   H_1   \big( d \rho^2 + dz^2\big)   ,
\end{gather*}
where
\begin{gather*}
   H_1  =  M_0 + c_1^2 M_1 + b_2^2 M_2 - b_2^2 c_1^2 M_3   , \\
   H_2  =  \frac{\mu_5}{\mu_3} \left( \frac{\mu_1}{\mu_7}   M_0
           - c_1^2 \frac{\rho^2}{\mu_1 \mu_7}   M_1 - b_2^2 \frac{\mu_1 \mu_7}{\rho^2}   M_2
           - b_2^2 c_1^2 \frac{\mu_7}{\mu_1}   M_3 \right)    , \\
   G_1  =  \frac{\rho^2 \mu_1 \mu_5}{\mu_2 \mu_4 \mu_6} \left( M_0
           - c_1^2 \frac{\rho^2}{\mu_1^2}   M_1 + b_2^2 M_2 + b_2^2 c_1^2 \frac{\rho^2}{\mu_1^2}   M_3
               \right)    , \\
   G_2  =  \frac{\mu_2 \mu_4 \mu_6}{\mu_3 \mu_7} \Big( M_0
           + c_1^2   M_1 - b_2^2 \frac{\mu_7^2}{\rho^2} M_2 + b_2^2 c_1^2 \frac{\mu_7^2}{\rho^2}   M_3
               \Big)    , \\
   \omega_1  =  b_2   \frac{\cR_{77}}{\mu_7}
           \left( \frac{\mu_1 \mu_5}{\mu_2 \mu_4 \mu_6} \right)^{1/2}
           \left( \sqrt{M_0 M_2} - c_1^2 \frac{\rho}{\mu_1} \sqrt{M_1 M_3} \right)     , \\
   \omega_2  =  c_1   \frac{\cR_{11}}{\mu_1} \left( \frac{\mu_2 \mu_4 \mu_6}{\rho^2 \mu_3 \mu_7}
               \right)^{1/2}
             \left( \sqrt{M_0 M_1} - b_2^2 \frac{\mu_7}{\rho} \sqrt{M_2 M_3} \right)     , \\
   J  =  b_2 c_1 \rho^2 \mu_1 \mu_2 \mu_3 \mu_4 \mu_5^2 \mu_6   (\mu_3-\mu_7)^2 (\mu_4-\mu_7)
         (\mu_5-\mu_7) (\mu_6-\mu_7) \\
\hphantom{J  =}{} \times\cR_{11} \cR_{12} \cR_{13} \cR_{14} \cR_{15}^2 \cR_{16} \cR_{17} \cR_{27} \cR_{77}   ,
\end{gather*}
and
\begin{gather*}
   M_0  =  \mu_4 \mu_5^3 \mu_6 \mu_7   (\mu_3-\mu_7)^4   \cR_{12}^2 \cR_{13}^2 \cR_{14}^2 \cR_{16}^2
           \cR_{17}^2 \cR_{27}^2   , \\
   M_1  =  \rho^2 \mu_1^2 \mu_2 \mu_3 \mu_4^2 \mu_5 \mu_6^2   (\mu_1-\mu_7)^2 (\mu_3-\mu_7)^4
             \cR_{15}^4 \cR_{17}^2 \cR_{27}^2     , \\
   M_2  =  \rho^4 \mu_1 \mu_2 \mu_3^2 \mu_5^2 \mu_7   (\mu_4-\mu_7)^2 (\mu_5-\mu_7)^2 (\mu_6-\mu_7)^2
             \cR_{12}^2 \cR_{13}^2 \cR_{14}^2 \cR_{16}^2      , \\
   M_3  =  \rho^4 \mu_1^3 \mu_2^2 \mu_3^3 \mu_4 \mu_6   (\mu_4-\mu_7)^2 (\mu_5-\mu_7)^2 (\mu_6-\mu_7)^2
             \cR_{15}^4 \cR_{17}^2     , \\
   \cP  =  \frac{\mu_2}{\mu_1 \mu_5^4 \mu_7 (\mu_3-\mu_7)^4}
           \frac{ \cR_{23} \cR_{25} \cR_{34} \cR_{35} \cR_{36} \cR_{45} \cR_{47} \cR_{56}
           \cR_{57} \cR_{67} }{ \cR_{12} \cR_{13} \cR_{14} \cR_{15}^2 \cR_{16} \cR_{17}
           \cR_{24}^2 \cR_{26}^2 \cR_{27} \cR_{37}^2 \cR_{46}^2 }   .
\end{gather*}
With obvious changes in notation, this is the ``bicycling'' black bi-ring solution
obtained in~\cite{Elva+Rodr08} (also see~\cite{Izumi08}), except for the fact that
we have a minus sign instead of a plus in the expressions for~$\omega_1$ and~$\omega_2$.

\section{Final remarks}
\label{sec:conclusions}
We presented a general formulation of binary Darboux-type transformations in the bidif\/ferential
calculus framework. Whenever a PDDE can be cast into the form (\ref{phi_eq}) or (\ref{g_eq}),
Theorem~\ref{thm:main} can be applied and it will typically generate
a large class of exact solutions. Meanwhile a~bidif\/ferential calculus formulation
is available for quite a number of integrable PDDEs.

We elaborated this general result for the case of the non-autonomous chiral model,
considerably extending previous results in \cite{DKMH11sigma}. We also presented
conditions that, imposed on the matrix data that determine the general class of solutions,
guarantee that the resulting solution of the non-autonomous chiral model is symmetric
(or Hermitian). If the solution is also real, then it is known to determine a Ricci-f\/lat metric,
i.e., a solution of the vacuum Einstein equations, dimensionally reduced to two dimensions.
We essentially solved the equations resulting from the assumptions in Theorem~\ref{thm:main}
in the case of a diagonal seed metric, though not yet the $\bV$-equations in suf\/f\/icient
generality if $\tbP$ is \emph{non}-diagonal (but see Examples~\ref{ex:Q_Jordan} and \ref{ex:TS}).
All this provides a working recipe to compute quite easily solutions of the vacuum
Einstein equations. In particular, in the four-dimensional case we recovered (multi-)
Kerr-NUT (in a dif\/ferent way than in \cite{DKMH11sigma}) and the
$\delta=2$ Tomimatsu--Sato solution. In the f\/ive-dimensional case we recovered
single and double Myers--Perry black holes, the ``black saturn'' and the ``bicycling
black ring'' solutions. The more general solutions still have to be explored.
In view of the complexity of the latter solutions, it is certainly an advantage to have now
an independent method at our disposal to derive, verify or generalize them.
Surely further important solutions of Einstein's equations in $D \geq 4$ space-time
dimensions can be recovered using this method and there is a chance to discover interesting
new solutions.
We concentrated on examples in the stationary case $\epsilon=1$, but developed the
formalism as well for the wave case $\epsilon=-1$ \cite{Beli+Zakh78,Beli+Verd01}. It is
not dif\/f\/icult to recover relevant examples in this case too.

The recipe to construct solutions of the non-autonomous chiral model and the
dimensionally reduced vacuum Einstein equations, obtained from Theorem~\ref{thm:main},
is~-- not surprisingly~-- a variant, a sort of matrix version, of the well-known method
of Belinski and Zakharov \cite{Beli+Zakh78,Beli+Sakh79,Beli+Verd01}.

One should look for suitable ways to spot physically relevant solutions within the plethora
of solutions. How are desired properties of solutions, like asymptotic
f\/latness, absence of naked singularities and proper axis conditions
encoded in the (matrix) data that
determine a solution? Here the rod structure analysis
\cite{Beli+Verd01,Empa+Real02PRD,Harm04,Chen+Teo10},
developed for the Belinski--Zakharov approach and frequently used, is of great help.

Section~\ref{sec:naCM} also paved the way toward a treatment of other reductions of the
non-autonomous chiral model, which, e.g., are relevant in the Einstein--Maxwell case and
supergravity theories.

In this work we only elaborated Theorem~\ref{thm:main}
for a particular example of an integrable equation in the bidif\/ferential calculus framework.
Although we already applied a more restricted version of it previously to several other integrable equations,
it will be worth to reconsider them and to also explore further equations, using the much more
general solution-generating tool we now have at our disposal. Furthermore, it should be
clarif\/ied whether, e.g., the examples in \cite{Sakh94,Sakh01,Sakh10} f\/it into the framework
of Theorem~\ref{thm:main}. We should also mention that Sylvester equations, like those that arise from
(\ref{preSylvester}), and more generally operator versions of them, are
ubiquitous in the theory of integrable systems. In particular, they are related
to a Riemann--Hilbert factorization problem~\cite{SakhL86} and they are at the roots of
Marchenko's operator approach~\cite{March88}.

\renewcommand{\theequation} {\Alph{section}.\arabic{equation}}
\renewcommand{\thesection} {\Alph{section}}

\makeatletter
\newcommand\appendix@section[1]{
  \refstepcounter{section}
  \orig@section*{Appendix \@Alph\c@section. #1}
  \addcontentsline{toc}{section}{Appendix \@Alph\c@section. #1}
}
\let\orig@section\section
\g@addto@macro\appendix{\let\section\appendix@section}
\makeatother

\begin{appendix}

\section{Addendum to Example~\ref{ex:f_for_diag_g}}
\label{app:f}
 From (\ref{frakf_eqs}) we f\/ind that
\[
   \mathfrak{f}[ \rho ] \propto \rho^{1/4}   , \qquad
   \mathfrak{f}[w^k] \propto \mathfrak{f}[w]^{k^2}   , \qquad
   \mathfrak{f}[w_1 w_2] \propto \mathfrak{f}[w_1]   \mathfrak{f}[w_2]   \mathfrak{F}[w_1,w_2]    ,
\]
where $\mathfrak{F}[w_1,w_2]$ has to solve
\begin{gather*}
 (\ln \mathfrak{F}[w_1,w_2])_\rho = \frac{\rho}{2} \big( (\ln w_1)_\rho   (\ln w_2)_\rho
                            - \epsilon   (\ln w_1)_z   (\ln w_2)_z \big)   , \\
  (\ln \mathfrak{F}[w_1,w_2])_z = \frac{\rho}{2} \big( (\ln w_1)_\rho   (\ln w_2)_z
                            + (\ln w_1)_z   (\ln w_2)_\rho \big)
\end{gather*}
(also see \cite{Izumi08}). It is easy to verify that
\[
           \mathfrak{F}[w_1 \cdots w_r , w_1' \cdots w_s']
  \propto  \prod_{i=1}^r \prod_{j=1}^s \mathfrak{F}[w_i , w_j']   .
\]
In particular, $\mathfrak{F}[w_1^k, w_2^l] \propto \mathfrak{F}[w_1,w_2]^{k l}$.
Furthermore, we have
\[
    \mathfrak{F}\big[\rho^k   w_1, \rho^l   w_2\big] \propto \rho^{kl/2}   w_1^{l/2} w_2^{k/2}
           \mathfrak{F}[w_1,w_2]   .
\]
It follows that
\begin{gather*}
 \mathfrak{f}[\rho^k   w] \propto \rho^{k^2/4} w^{k/2} \mathfrak{f}[w]   ,   \qquad
 \mathfrak{f}[\tilde{\mu}_1 \cdots \tilde{\mu}_r] \propto
       \left(\prod_{k=1}^r \mathfrak{f}[\tilde{\mu}_k] \right)
       \left(\prod_{i < j} \mathfrak{F}[\tilde{\mu}_i , \tilde{\mu}_j] \right)   ,
\end{gather*}
and
\[
  \mathfrak{f} \left[ \frac{\tilde{\mu}_1 \cdots \tilde{\mu}_r}{\tilde{\mu}'_1 \cdots \tilde{\mu}'_s} \right]
  \propto \frac{ \mathfrak{f} [ \tilde{\mu}_1 \cdots \tilde{\mu}_r ]
       \mathfrak{f} [ \tilde{\mu}'_1 \cdots \tilde{\mu}'_r ] }
     { \prod\limits_i^r \prod\limits_j^s \mathfrak{F}[\tilde{\mu}_i, \tilde{\mu}'_j] }   ,
\]
from which the main result in Example~\ref{ex:f_for_diag_g} is easily deduced.

\section{Some proofs}
\label{app:proofs}

{\bf Proof of Proposition~\ref{prop:f_sol}.}
Using $(\ln \det Y)_\rho = \operatorname{tr}(Y_\rho Y^{-1})$ for an invertible (and
dif\/ferentiable) matrix function $Y$, we obtain
\begin{gather*}
  (\ln f)_\rho  = (\ln f_0)_\rho + \operatorname{tr}\Big({-}\rho^{-1} \bI + \tbX_\rho \tbX^{-1}
                   + \tbP_\rho \tbP^{-1} + \tbQ_\rho \tbQ^{-1}
                   - \epsilon \tbP_\rho \tbP \big(1+\epsilon   \tbP^2\big)^{-1} \\
\hphantom{(\ln f)_\rho  =}{}
 - \epsilon \tbQ_\rho \tbQ \big(1+\epsilon   \tbQ^2\big)^{-1} \Big)   , \\
  (\ln f)_z  =  (\ln f_0)_z + \operatorname{tr}\Big( \tbX_z \tbX^{-1} + \tbP_z \tbP^{-1}
                + \tbQ_z \tbQ^{-1}
                - \epsilon \tbP_z \tbP \big(1+\epsilon   \tbP^2\big)^{-1}  \\
\hphantom{(\ln f)_z  =}{}
   - \epsilon \tbQ_z \tbQ \big(1+\epsilon   \tbQ^2\big)^{-1} \Big)   ,
\end{gather*}
for the expression of $f$ in Proposition~\ref{prop:f_sol}. In order to verify that
(\ref{f_eqs}) holds, we have to show that these expressions equal the corresponding
right hand sides of (\ref{f_eqs}), evaluated with
\begin{gather*}
 \cU  =  -\rho   \tphi_z = -\rho   \big[\tphi_{0,z} + \big(\bU \tbX^{-1} \bV\big)_z \big] \\
 \hphantom{\cU  =}{}
     =  \cU_0 - \rho   \big( \bU_z \tbX^{-1} \bV - \bU \tbX^{-1} \tbX_z \tbX^{-1} \bV
         + \bU \tbX^{-1} \bV_z \big)   , \\
 \cV  =  \cV_0 + \epsilon   \big[ \rho   \big( \bU_\rho \tbX^{-1} \bV - \bU \tbX^{-1} \tbX_\rho \tbX^{-1} \bV
         + \bU \tbX^{-1} \bV_\rho \big) + \bU \tbX^{-1} \bV \big]   ,
\end{gather*}
where $\cU_0 = -\rho   \tphi_{0,z}$ and
$\cV_0 = \epsilon   (\rho   \tphi_{0,\rho} + \tphi_0)$.\footnote{This is quite
a tour de force and a more elegant proof would be desirable.}
We work on the right hand sides of (\ref{f_eqs}) and, as intermediate steps, we consider
$\operatorname{tr}(\cU^2 - \cU_0^2)$, $\operatorname{tr}(\cV^2 - \cV_0^2)$
and $\operatorname{tr}(\cU \cV - \cU_0 \cV_0)$ separately.
First we eliminate derivatives of $\phi_0$ with the help of (\ref{naCM_U_eqs}) and
(\ref{naCM_V_eqs}). Then we eliminate $\bV \bU$ using the Sylvester equation,
and $\bV_z   \bU$, $\bV_\rho   \bU$ via~(\ref{naCM_X_eqs}).
Rewriting (\ref{naCM_X_eqs}) with the help of (\ref{naCM_Sylv}), we obtain a
version that allows us to also replace all occurencies of $\bV \bU_z$ and $\bV \bU_\rho$.
Several times one has to exploit the cyclicity of the trace.
Finally we use
\[
    \tbP_\rho = \rho^{-1} \tbP \big(\bI - \epsilon \tbP^2\big)   \big(\bI + \epsilon \tbP^2\big)^{-1}   , \qquad
    \tbP_z = 2 \epsilon   \rho^{-1} \tbP^2 \big(\bI + \epsilon \tbP^2\big)^{-1}
\]
(which follows from (\ref{tbP_eqs})), and the corresponding equations for $\tbQ$,
to show that (\ref{f_eqs}) holds.

\medskip

\noindent
{\bf Proof of Proposition~\ref{prop:f_transform}.}
Using $\hat{\cU} = \cU + \rho   (\ln w)_\rho   I$ and $\hat{\cV} = \cV + \rho   (\ln w)_z   I$,
we have
\[
 \operatorname{tr}\big(\hat{\cU}^2 - \cU^2\big) =  2 \rho   (\ln w)_\rho   \operatorname{tr} \cU
                              + m   \rho^2 (\ln w)_\rho^2
                         = 2 \rho^2   (\ln w)_\rho   (\ln \det g)_\rho
                              + m   \rho^2 (\ln w)_\rho^2   ,
\]
and a corresponding expression for $\operatorname{tr} (\hat{\cV}^2 - \cV^2)$.  Furthermore,
\[
    \operatorname{tr}\big( \hat{\cU} \hat{\cV} - \cU \cV \big)
  = \rho^2 \big( (\ln w)_\rho (\ln \det g)_z + (\ln w)_z (\ln \det g)_\rho + m   (\ln w)_\rho (\ln w)_z \big)   .
\]
With the help of (\ref{f_eqs}) we obtain
\begin{gather*}
    \left( \ln \frac{\hat{f}}{f} \right)_\rho
   =  \frac{1}{4 \rho} \operatorname{tr}\big( \hat{\cU}^2 - \cU^2  - \epsilon   \big(\hat{\cV}^2 - \cV^2\big) \big) \\
\hphantom{\left( \ln \frac{\hat{f}}{f} \right)_\rho}{}
  =  \frac{\rho}{2} \big( (\ln w)_\rho   (\ln \det g)_\rho - \epsilon   (\ln w)_z   (\ln \det g)_z \big)
      + m   \frac{\rho}{4} \left( (\ln w)_\rho^2 - \epsilon   (\ln w)_z^2 \right)   , \\
    \left( \ln \frac{\hat{f}}{f} \right)_z
   =  \frac{1}{2 \rho} \operatorname{tr}\big( \hat{\cU} \hat{\cV} - \cU \cV \big) \\
\hphantom{\left( \ln \frac{\hat{f}}{f} \right)_z}{}
  =  \frac{\rho}{2} \big( (\ln w)_\rho (\ln \det g)_z
         + (\ln w)_z (\ln \det g)_\rho \big) + m   \frac{\rho}{2} (\ln w)_\rho (\ln w)_z   .
\end{gather*}
Next we use
\begin{gather*}
    (\ln(\rho   f_{(w)}))_\rho = \frac{\rho}{4} \big( (\ln w)_\rho^2 - \epsilon   (\ln w)_z^2 \big)   , \qquad
    (\ln(\rho   f_{(w)}))_z = \frac{\rho}{2} (\ln w)_\rho   (\ln w)_z   ,
\end{gather*}
and the corresponding equations for $f_{(\det g)}$ and $f_{(w   \det g)}$, to deduce that
\begin{gather*}
    \left( \ln \frac{f_{(w   \det g)}}{\rho   f_{(w)}   f_{(\det g)} } \right)_\rho
  =  \frac{\rho}{2} \big( (\ln w)_\rho   (\ln \det g)_\rho - \epsilon   (\ln w)_z   (\ln \det g)_z \big)   , \\
    \left( \ln \frac{f_{(w   \det g)}}{\rho   f_{(w)}   f_{(\det g)} } \right)_z
   =  \frac{\rho}{2} \big( (\ln w)_\rho   (\ln \det g)_z + (\ln w)_z   (\ln \det g)_\rho \big)   .
\end{gather*}
Inserting the last expressions in our previous results, we obtain (\ref{f_tranform}) by integration.
Let us now assume that $\det \hat{g} = -\epsilon   \rho^2$, and thus $\det g = -\epsilon   \rho^2 w^{-m}$.
With
\[
     f_{(\rho^2 w^k)} \propto w^k (\rho   f_{(w)})^{k^2}
\]
(see Example~\ref{ex:f_for_diag_g}), (\ref{f_tranform_special}) results from (\ref{f_tranform}).

\end{appendix}

\pdfbookmark[1]{References}{ref}
\LastPageEnding


\begin{thebibliography}{99}
\footnotesize\itemsep=0pt

\bibitem{Arsi+Lore11}
Arsie A., Lorenzoni P., $F$-manifolds with eventual identities, bidif\/ferential
  calculus and twisted Lenard--Magri chains, \href{http://dx.doi.org/10.1093/imrn/rns172}{\textit{Int. Math. Res. Not.}}, {t}o
  appear, \href{http://arxiv.org/abs/1110.2461}{arXiv:1110.2461}.

\bibitem{Beli+Verd01}
Belinski V.A., Verdaguer E., Gravitational solitons, \href{http://dx.doi.org/10.1017/CBO9780511535253}{\textit{Cambridge Monographs on
  Mathematical Physics}}, Cambridge University Press, Cambridge, 2001.

\bibitem{Beli+Sakh79}
Belinski{\u\i} V.A., Sakharov V.E., Stationary gravitational solitons with
  axial symmetry, \textit{Sov. Phys. JETP} \textbf{50} (1979), 1--9.

\bibitem{Beli+Zakh78}
Belinski{\u\i} V.A., Zakharov V.E., Integration of the Einstein equations by
  means of the inverse scattering problem technique and construction of exact
  soliton solutions, \textit{Sov. Phys. JETP} \textbf{48} (1978), 985--994.

\bibitem{BMG88}
Breitenlohner P., Maison D., Gibbons G., {$4$}-dimensional black holes from
  {K}aluza--{K}lein theories, \href{http://dx.doi.org/10.1007/BF01217967}{\textit{Comm. Math. Phys.}} \textbf{120} (1988),
  295--333.

\bibitem{Cama+Cari03}
Camacaro J., Cari{\~n}ena J., Alternative Lie algebroid structures and
  bi-dif\/ferential calculi, in Applied Dif\/ferential Geometry and Mechanics,
  Editors W.~Sarlet, F.~Cantrijn, University of Gent, 2003, 1--20.

\bibitem{Chav05}
Chavchanidze G., Non-{N}oether symmetries in {H}amiltonian dynamical systems,
  \textit{Mem. Differential Equations Math. Phys.} \textbf{36} (2005), 81--134,
  \href{http://arxiv.org/abs/math-ph/0405003}{math-ph/0405003}.

\bibitem{CHT11}
Chen Y., Hong K., Teo E., Unbalanced Pomeransky--Sen'kov black ring,
  \href{http://dx.doi.org/10.1103/PhysRevD.84.084030}{\textit{Phys. Rev.~D}} \textbf{84} (2011), 084030, 11~pages,
  \href{http://arxiv.org/abs/1108.1849}{arXiv:1108.1849}.

\bibitem{Chen+Teo10}
Chen Y., Teo E., Rod-structure classif\/ication of gravitational instantons with
  {$U(1)\times U(1)$} isometry, \href{http://dx.doi.org/10.1016/j.nuclphysb.2010.05.017}{\textit{Nuclear Phys.~B}} \textbf{838} (2010),
  207--237, \href{http://arxiv.org/abs/1004.2750}{arXiv:1004.2750}.

\bibitem{Chru+Cort10}
Chru\'{s}ciel P., Cortier J., Maximal analytic extensions of the Emparan--Reall
  black ring, \href{http://dx.doi.org/10.1088/1742-6596/229/1/012030}{\textit{J.~Phys. Conf. Ser.}} \textbf{229} (2010), 012030,
  4~pages, \href{http://arxiv.org/abs/0807.2309}{arXiv:0807.2309}.

\bibitem{CES10}
Chru{\'s}ciel P.T., Eckstein M., Szybka S.J., On smoothness of black saturns,
  \href{http://dx.doi.org/10.1007/JHEP11(2010)048}{\textit{J.~High Energy Phys.}} \textbf{2010} (2010), no.~11, 048, 39~pages,
  \href{http://arxiv.org/abs/1007.3668}{arXiv:1007.3668}.

\bibitem{Cies09}
Cie{\'s}li{\'n}ski J.L., Algebraic construction of the {D}arboux matrix
  revisited, \href{http://dx.doi.org/10.1088/1751-8113/42/40/404003}{\textit{J.~Phys.~A: Math. Theor.}} \textbf{42} (2009), 404003,
  40~pages, \href{http://arxiv.org/abs/0904.3987}{arXiv:0904.3987}.

\bibitem{CST00b}
Crampin M., Sarlet W., Thompson G., Bi-dif\/ferential calculi, bi-{H}amiltonian
  systems and conformal {K}illing tensors, \href{http://dx.doi.org/10.1088/0305-4470/33/48/313}{\textit{J.~Phys.~A: Math. Gen.}}
  \textbf{33} (2000), 8755--8770.

\bibitem{deSouza+Bhatt81}
de~Souza E., Bhattacharyya S.P., Controllability, observability and the
  solution of {$AX-XB=C$}, \href{http://dx.doi.org/10.1016/0024-3795(81)90301-3}{\textit{Linear Algebra Appl.}} \textbf{39} (1981),
  167--188.

\bibitem{DKMH11sigma}
Dimakis A., Kanning N., M{\"u}ller-Hoissen F., The non-autonomous chiral model
  and the {E}rnst equation of general relativity in the bidif\/ferential calculus
  framework, \href{http://dx.doi.org/10.3842/SIGMA.2011.118}{\textit{SIGMA}} \textbf{7} (2011), 118, 27~pages,
  \href{http://arxiv.org/abs/1106.4122}{arXiv:1106.4122}.

\bibitem{DMH00a}
Dimakis A., M{\"u}ller-Hoissen F., Bi-dif\/ferential calculi and integrable
  models, \href{http://dx.doi.org/10.1088/0305-4470/33/5/311}{\textit{J.~Phys.~A: Math. Gen.}} \textbf{33} (2000), 957--974,
  \href{http://arxiv.org/abs/math-ph/9908015}{math-ph/9908015}.

\bibitem{DMH00e}
Dimakis A., M{\"u}ller-Hoissen F., Bicomplexes and integrable models,
  \href{http://dx.doi.org/10.1088/0305-4470/33/37/310}{\textit{J.~Phys.~A: Math. Gen.}} \textbf{33} (2000), 6579--6591,
  \href{http://arxiv.org/abs/nlin.SI/0006029}{nlin.SI/0006029}.

\bibitem{DMH10AKNS}
Dimakis A., M{\"u}ller-Hoissen F., Bidif\/ferential calculus approach to {AKNS}
  hierarchies and their solutions, \href{http://dx.doi.org/10.3842/SIGMA.2010.055}{\textit{SIGMA}} \textbf{6} (2010), 055,
  27~pages, \href{http://arxiv.org/abs/1004.1627}{arXiv:1004.1627}.

\bibitem{DMH08bidiff}
Dimakis A., M{\"u}ller-Hoissen F., Bidif\/ferential graded algebras and
  integrable systems, \textit{Discrete Contin. Dyn. Syst.}  (2009), suppl.,
  208--219, \href{http://arxiv.org/abs/0805.4553}{arXiv:0805.4553}.


\bibitem{DMH10NLS}
Dimakis A., M{\"u}ller-Hoissen F., Solutions of matrix {NLS} systems and their
  discretizations: a unif\/ied treatment, \href{http://dx.doi.org/10.1088/0266-5611/26/9/095007}{\textit{Inverse Problems}} \textbf{26}
  (2010), 095007, 55~pages, \href{http://arxiv.org/abs/1001.0133}{arXiv:1001.0133}.

\bibitem{Econ+Tsou89}
Economou A., Tsoubelis D., Multiple-soliton solutions of {E}instein's
  equations, \href{http://dx.doi.org/10.1063/1.528597}{\textit{J.~Math. Phys.}} \textbf{30} (1989), 1562--1569.

\bibitem{Elva+Figu07}
Elvang H., Figueras P., Black saturn, \href{http://dx.doi.org/10.1088/1126-6708/2007/05/050}{\textit{J.~High Energy Phys.}}
  \textbf{2007} (2007), no.~5, 050, 48 pages, \mbox{\href{http://arxiv.org/abs/hep-th/0701035}{hep-th/0701035}}.

\bibitem{Elva+Rodr08}
Elvang H., Rodriguez M.J., Bicycling black rings, \href{http://dx.doi.org/10.1088/1126-6708/2008/04/045}{\textit{J.~High Energy Phys.}}
  \textbf{2008} (2008), no.~4, 045, 30~pages, \href{http://arxiv.org/abs/0712.2425}{arXiv:0712.2425}.

\bibitem{Empa+Real02PRL}
Emparan R., Reall H.S., A rotating black ring solution in f\/ive dimensions,
  \href{http://dx.doi.org/10.1103/PhysRevLett.88.101101}{\textit{Phys. Rev. Lett.}} \textbf{88} (2002), 101101, 4~pages,
  \href{http://arxiv.org/abs/hep-th/0110260}{hep-th/0110260}.

\bibitem{Empa+Real08}
Emparan R., Reall H.S., Black holes in higher dimensions, \textit{Living Rev.
  Relativ.} \textbf{11} (2008), 6, 87~pages, \href{http://arxiv.org/abs/0801.3471}{arXiv:0801.3471}.

\bibitem{Empa+Real06}
Emparan R., Reall H.S., Black rings, \href{http://dx.doi.org/10.1088/0264-9381/23/20/R01}{\textit{Classical Quantum Gravity}}
  \textbf{23} (2006), R169--R197, \href{http://arxiv.org/abs/hep-th/0608012}{hep-th/0608012}.


\bibitem{Empa+Real02PRD}
Emparan R., Reall H.S., Generalized {W}eyl solutions, \href{http://dx.doi.org/10.1103/PhysRevD.65.084025}{\textit{Phys. Rev.~D}}
  \textbf{65} (2002), 084025, 26~pages, \mbox{\href{http://arxiv.org/abs/hep-th/0110258}{hep-th/0110258}}.

\bibitem{Evsl+Kris09}
Evslin J., Krishnan C., The black di-ring: an inverse scattering construction,
  \href{http://dx.doi.org/10.1088/0264-9381/26/12/125018}{\textit{Classical Quantum Gravity}} \textbf{26} (2009), 125018, 13~pages,
  \href{http://arxiv.org/abs/0706.1231}{arXiv:0706.1231}.

\bibitem{Figueras05}
Figueras P., A black ring with a rotating 2-sphere, \href{http://dx.doi.org/10.1088/1126-6708/2005/07/039}{\textit{J.~High Energy
  Phys.}} \textbf{2005} (2005), no.~7, 039, 9 pages, \href{http://arxiv.org/abs/hep-th/0505244}{hep-th/0505244}.

\bibitem{FJRV10}
Figueras P., Jamsin E., Rocha J.V., Virmani A., Integrability of
  f\/ive-dimensional minimal supergravity and charged rotating black holes,
  \href{http://dx.doi.org/10.1088/0264-9381/27/13/135011}{\textit{Classical Quantum Gravity}} \textbf{27} (2010), 135011, 37~pages,
  \href{http://arxiv.org/abs/0912.3199}{arXiv:0912.3199}.

\bibitem{Froe+Nije56}
Fr\"olicher A., Nijenhuis A., Theory of vector-valued dif\/ferential forms.
  I.~Derivations in the graded ring of dif\/ferential forms, \textit{Proc.
  Koninkl. Ned. Acad. Wetensch. Ser.~A} \textbf{59} (1956), 338--359.

\bibitem{Frolov07}
Frolov V.P., Goswami R., Surface geometry of 5{D} black holes and black rings,
  \href{http://dx.doi.org/10.1103/PhysRevD.75.124001}{\textit{Phys. Rev.~D}} \textbf{75} (2007), 124001, 11~pages,
  \href{http://arxiv.org/abs/gr-qc/0612033}{gr-qc/0612033}.

\bibitem{Gaun+Guto05}
Gauntlett J.P., Gutowski J.B., Concentric black rings, \href{http://dx.doi.org/10.1103/PhysRevD.71.025013}{\textit{Phys. Rev.~D}}
  \textbf{71} (2005), 025013, 7~pages, \mbox{\href{http://arxiv.org/abs/hep-th/0408010}{hep-th/0408010}}.

\bibitem{Griff91}
Grif\/f\/iths J.B., Colliding plane waves in general relativity, \textit{Oxford
  Mathematical Monographs, Oxford Science Publications}, The Clarendon Press,
  Oxford University Press, New York, 1991.

\bibitem{Gu+Zhou05}
Gu C., Hu H., Zhou Z., Darboux transformations in integrable systems. Theory
  and their applications to geometry, \textit{Mathematical Physics Studies},
  Vol.~26, Springer, Dordrecht, 2005.

\bibitem{Harm04}
Harmark T., Stationary and axisymmetric solutions of higher-dimensional general
  relativity, \href{http://dx.doi.org/10.1103/PhysRevD.70.124002}{\textit{Phys. Rev.~D}} \textbf{70} (2004), 124002, 25~pages,
  \href{http://arxiv.org/abs/hep-th/0408141}{hep-th/0408141}.

\bibitem{Hartwig72}
Hartwig R.E., Resultants and the solution of {$AX-XB=-C$}, \href{http://dx.doi.org/10.1137/0123012}{\textit{SIAM~J.
  Appl. Math.}} \textbf{23} (1972), 104--117.

\bibitem{Hearon77}
Hearon J.Z., Nonsingular solutions of {$TA-BT=C$}, \href{http://dx.doi.org/10.1016/0024-3795(77)90019-2}{\textit{Linear Algebra
  Appl.}} \textbf{16} (1977), 57--63.

\bibitem{HRZC08}
Herdeiro C., Rebelo C., Zilh\~{a}o M., Costa M., A double Myers--Perry black
  hole in f\/ive dimensions, \href{http://dx.doi.org/10.1088/1126-6708/2008/07/009}{\textit{J.~High Energy Phys.}} \textbf{2008} (2008),
  no.~7, 009, 24~pages, \href{http://arxiv.org/abs/0805.1206}{arXiv:0805.1206}.

\bibitem{Holl+Yaza08}
Hollands S., Yazadjiev S., Uniqueness theorem for 5-dimensional black holes
  with two axial {K}illing f\/ields, \href{http://dx.doi.org/10.1007/s00220-008-0516-3}{\textit{Comm. Math. Phys.}} \textbf{283}
  (2008), 749--768, \href{http://arxiv.org/abs/0707.2775}{arXiv:0707.2775}.

\bibitem{Hong+Teo03}
Hong K., Teo E., A new form of the {$C$}-metric, \href{http://dx.doi.org/10.1088/0264-9381/20/14/321}{\textit{Classical Quantum
  Gravity}} \textbf{20} (2003), 3269--3277, \mbox{\href{http://arxiv.org/abs/gr-qc/0305089}{gr-qc/0305089}}.

\bibitem{Horowitz12}
Horowitz G. (Editor), Black holes in higher dimensions, Cambridge University
  Press, Cambridge, 2012.

\bibitem{Hoski09}
Hoskisson J., Explorations of four and f\/ive dimensional black hole spacetimes,
  Ph.D. thesis, Durham University, 2009, available at
  \url{http://etheses.dur.ac.uk/2115/}.

\bibitem{Hu+Cheng06}
Hu Q., Cheng D., The polynomial solution to the {S}ylvester matrix equation,
  \href{http://dx.doi.org/10.1016/j.aml.2005.09.005}{\textit{Appl. Math. Lett.}} \textbf{19} (2006), 859--864.

\bibitem{IIM11}
Iguchi H., Izumi K., Mishima T., Systematic solution-generation of
  f\/ive-dimensional black holes, \href{http://dx.doi.org/10.1143/PTPS.189.93}{\textit{Progr. Theoret. Phys. Suppl.}}
  \textbf{189} (2011), 93--125, \href{http://arxiv.org/abs/1106.0387}{arXiv:1106.0387}.

\bibitem{Iguchi+Mishima06}
Iguchi H., Mishima T., Solitonic generation of vacuum solutions in
  f\/ive-dimensional general relativity, \href{http://dx.doi.org/10.1103/PhysRevD.74.024029}{\textit{Phys. Rev.~D}} \textbf{74}
  (2006), 024029, 17~pages, \href{http://arxiv.org/abs/hep-th/0605090}{hep-th/0605090}.

\bibitem{Izumi08}
Izumi K., Orthogonal black di-ring solution, \href{http://dx.doi.org/10.1143/PTP.119.757}{\textit{Progr. Theoret. Phys.}}
  \textbf{119} (2008), 757--774, \href{http://arxiv.org/abs/0712.0902}{arXiv:0712.0902}.

\bibitem{Klein+Richter05}
Klein C., Richter O., Ernst equation and {R}iemann surfaces. Analytical and
  numerical methods, \textit{Lecture Notes in Physics}, Vol.~685,
  Springer-Verlag, Berlin, 2005.

\bibitem{Koda+Hiki03}
Kodama H., Hikida W., Global structure of the {Z}ipoy--{V}oorhees--{W}eyl
  spacetime and the {$\delta=2$} {T}omimatsu--{S}ato spacetime,
  \href{http://dx.doi.org/10.1088/0264-9381/20/23/011}{\textit{Classical Quantum Gravity}} \textbf{20} (2003), 5121--5140,
  \href{http://arxiv.org/abs/gr-qc/0304064}{gr-qc/0304064}.

\bibitem{Kramer+Neug80}
Kramer D., Neugebauer G., The superposition of two {K}err solutions,
  \href{http://dx.doi.org/10.1016/0375-9601(80)90556-3}{\textit{Phys. Lett.~A}} \textbf{75} (1980), 259--261.

\bibitem{Lorenzoni06}
Lorenzoni P., Flat bidif\/ferential ideals and semi-{H}amiltonian {PDE}s,
  \href{http://dx.doi.org/10.1088/0305-4470/39/44/006}{\textit{J.~Phys.~A: Math. Gen.}} \textbf{39} (2006), 13701--13715,
  \href{http://arxiv.org/abs/nlin.SI/0604053}{nlin.SI/0604053}.

\bibitem{Lore+Magri05}
Lorenzoni P., Magri F., A cohomological construction of integrable hierarchies
  of hydrodynamic type, \href{http://dx.doi.org/10.1155/IMRN.2005.2087}{\textit{Int. Math. Res. Not.}} \textbf{2005} (2005),
  no.~34, 2087--2100, \href{http://arxiv.org/abs/nlin.SI/0504064}{nlin.SI/0504064}.

\bibitem{March88}
Marchenko V.A., Nonlinear equations and operator algebras, \href{http://dx.doi.org/10.1007/978-94-009-2887-9}{\textit{Mathematics
  and its Applications (Soviet Series)}}, Vol.~17, D.~Reidel Publishing Co.,
  Dordrecht, 1988.

\bibitem{Matv+Sall91}
Matveev V.B., Salle M.A., Darboux transformations and solitons, \textit{Springer Series
  in Nonlinear Dynamics}, Springer-Verlag, Berlin, 1991.

\bibitem{Mish+Iguc06}
Mishima T., Iguchi H., New axisymmetric stationary solutions of
  f\/ive-dimensional vacuum {E}instein equations with asymptotic f\/latness,
  \href{http://dx.doi.org/10.1103/PhysRevD.73.044030}{\textit{Phys. Rev.~D}} \textbf{73} (2006), 044030, 6~pages,
  \href{http://arxiv.org/abs/hep-th/0504018}{hep-th/0504018}.

\bibitem{Myers12}
Myers R.C., Myers--Perry black holes, in Black Holes in Higher Dimensions,
  Editor G.~Horowitz, Cambridge University Press, Cambridge, 2012, 101--133,
  \href{http://arxiv.org/abs/1111.1903}{arXiv:1111.1903}.

\bibitem{Myers+Perry86}
Myers R.C., Perry M.J., Black holes in higher-dimensional space-times,
  \href{http://dx.doi.org/10.1016/0003-4916(86)90186-7}{\textit{Ann. Physics}} \textbf{172} (1986), 304--347.

\bibitem{Naka83}
Nakamura Y., Symmetries of stationary axially symmetric vacuum {E}instein
  equations and the new family of exact solutions, \href{http://dx.doi.org/10.1063/1.525735}{\textit{J.~Math. Phys.}}
  \textbf{24} (1983), 606--609.

\bibitem{NGO00}
Nimmo J.J.C., Gilson C., Ohta Y., Applications of {D}arboux transformations to the
  self-dual {Y}ang--{M}ills equations, \href{http://dx.doi.org/10.1007/BF02551200}{\textit{Theoret. and Math. Phys.}}
  \textbf{122} (2000), 239--246.

\bibitem{Pomer06}
Pomeransky A.A., Complete integrability of higher-dimensional {E}instein
  equations with additional symmetry and rotating black holes, \href{http://dx.doi.org/10.1103/PhysRevD.73.044004}{\textit{Phys.
  Rev.~D}} \textbf{73} (2006), 044004, 5~pages, \href{http://arxiv.org/abs/hep-th/0507250}{hep-th/0507250}.

\bibitem{Pomer+Senkov06}
Pomeransky A.A., Sen'kov R.A., Black ring with two angular momenta,
  \href{http://arxiv.org/abs/hep-th/0612005}{hep-th/0612005}.

\bibitem{Roge+Schi02}
Rogers C., Schief W.K., B\"acklund and {D}arboux transformations. Geometry and
  modern applications in soliton theory, \href{http://dx.doi.org/10.1017/CBO9780511606359}{\textit{Cambridge Texts in Applied
  Mathematics}}, Cambridge University Press, Cambridge, 2002.

\bibitem{Sakh94}
Sakhnovich A.L., Dressing procedure for solutions of nonlinear equations and
  the method of operator identities, \href{http://dx.doi.org/10.1088/0266-5611/10/3/013}{\textit{Inverse Problems}} \textbf{10}
  (1994), 699--710.

\bibitem{Sakh01}
Sakhnovich A.L., Generalized {B}\"acklund--{D}arboux transformation: spectral
  properties and nonlinear equations, \href{http://dx.doi.org/10.1006/jmaa.2001.7577}{\textit{J.~Math. Anal. Appl.}}
  \textbf{262} (2001), 274--306.

\bibitem{Sakh10}
Sakhnovich A.L., On the {GBDT} version of the {B}\"acklund--{D}arboux
  transformation and its applications to linear and nonlinear equations and
  {W}eyl theory, \href{http://dx.doi.org/10.1051/mmnp/20105415}{\textit{Math. Model. Nat. Phenom.}} \textbf{5} (2010),
  340--389, \href{http://arxiv.org/abs/0909.1537}{arXiv:0909.1537}.

\bibitem{SakhL86}
Sakhnovich L.A., Problems of factorization and operator identities,
  \href{http://dx.doi.org/10.1070/RM1986v041n01ABEH003200}{\textit{Russ. Math. Surv.}} \textbf{41} (1986), no.~1, 1--64.

\bibitem{SVV02}
Sparano G., Vilasi G., Vinogradov A.M., Vacuum {E}instein metrics with
  bidimensional {K}illing leaves. {I}.~{L}ocal aspects, \href{http://dx.doi.org/10.1016/S0926-2245(01)00062-6}{\textit{Differential
  Geom. Appl.}} \textbf{16} (2002), 95--120, \href{http://arxiv.org/abs/gr-qc/0301020}{gr-qc/0301020}.

\bibitem{SKMHH03}
Stephani H., Kramer D., MacCallum M., Hoenselaers C., Herlt E., Exact solutions
  of {E}instein's f\/ield equations, 2nd ed., \href{http://dx.doi.org/10.1017/CBO9780511535185}{\textit{Cambridge Monographs on
  Mathematical Physics}}, Cambridge University Press, Cambridge, 2003.

\bibitem{Szyb11}
Szybka S.J., Stable causality of {B}lack {S}aturns, \href{http://dx.doi.org/10.1007/JHEP05(2011)052}{\textit{J.~High Energy
  Phys.}} \textbf{2011} (2011), no.~5, 052, 9~pages, \href{http://arxiv.org/abs/1102.3942}{arXiv:1102.3942}.

\bibitem{Tan+Teo03}
Tan H.S., Teo E., Multi-black-hole solutions in f\/ive dimensions, \href{http://dx.doi.org/10.1103/PhysRevD.68.044021}{\textit{Phys.
  Rev.~D}} \textbf{68} (2003), 044021, 11~pages, \href{http://arxiv.org/abs/hep-th/0306044}{hep-th/0306044}.

\bibitem{Tangh63}
Tangherlini F.R., Schwarzschild f\/ield in {$n$} dimensions and the
  dimensionality of space problem, \href{http://dx.doi.org/10.1007/BF02784569}{\textit{Nuovo Cimento}} \textbf{27} (1963),
  636--651.

\bibitem{Tomi+Sato72}
Tomimatsu A., Sato H., New exact solution for the gravitational f\/ield of a
  spinning mass, \href{http://dx.doi.org/10.1103/PhysRevLett.29.1344}{\textit{Phys. Rev. Lett.}} \textbf{29} (1972), 1344--1345.

\bibitem{Tomi+Ishi11}
Tomizawa S., Ishihara H., Exact solutions of higher dimensional black holes,
  \href{http://arxiv.org/abs/1104.1468}{arXiv:1104.1468}.

\bibitem{TMY06}
Tomizawa S., Morisawa Y., Yasui Y., Vacuum solutions of f\/ive dimensional
  {E}instein equations generated by inverse scattering method, \href{http://dx.doi.org/10.1103/PhysRevD.73.064009}{\textit{Phys.
  Rev.~D}} \textbf{73} (2006), 064009, 8~pages, \href{http://arxiv.org/abs/hep-th/0512252}{hep-th/0512252}.

\bibitem{Tomi+Noza06}
Tomizawa S., Nozawa M., Vacuum solutions of f\/ive dimensional {E}instein
  equations generated by inverse scattering method. {II}.~{P}roduction of the
  black ring solution, \href{http://dx.doi.org/10.1103/PhysRevD.73.124034}{\textit{Phys. Rev.~D}} \textbf{73} (2006), 124034,
  10~pages, \href{http://arxiv.org/abs/hep-th/0604067}{hep-th/0604067}.

\bibitem{TUS04}
Tomizawa S., Uchida Y., Shiromizu M., Twist of stationary black hole or ring in
  f\/ive dimensions, \href{http://dx.doi.org/10.1103/PhysRevD.70.064020}{\textit{Phys. Rev.~D}} \textbf{70} (2004), 064020, 5~pages,
  \href{http://arxiv.org/abs/gr-qc/0405134}{gr-qc/0405134}.

\bibitem{Verd93}
Verdaguer E., Soliton solutions in spacetime with two spacelike {K}illing
  f\/ields, \href{http://dx.doi.org/10.1016/0370-1573(93)90139-5}{\textit{Phys. Rep.}} \textbf{229} (1993), 1--80.

\bibitem{Yazad08}
Yazadjiev S.S., 5{D} {E}instein--{M}axwell solitons and concentric rotating
  dipole black rings, \href{http://dx.doi.org/10.1103/PhysRevD.78.064032}{\textit{Phys. Rev.~D}} \textbf{78} (2008), 064032,
  11~pages, \href{http://arxiv.org/abs/0805.1600}{arXiv:0805.1600}.

\bibitem{Yazad07}
Yazadjiev S.S., Black saturn with a dipole ring, \href{http://dx.doi.org/10.1103/PhysRevD.76.064011}{\textit{Phys. Rev.~D}}
  \textbf{76} (2007), 064011, 8~pages, \href{http://arxiv.org/abs/0705.1840}{arXiv:0705.1840}.

\bibitem{Yazad06}
Yazadjiev S.S., Completely integrable sector in 5{D} {E}instein--{M}axwell
  gravity and derivation of the dipole black ring solutions, \href{http://dx.doi.org/10.1103/PhysRevD.73.104007}{\textit{Phys.
  Rev.~D}} \textbf{73} (2006), 104007, 7~pages, \href{http://arxiv.org/abs/hep-th/0602116}{hep-th/0602116}.




\end{thebibliography}
\end{document}